%% file: BiCoPO5_magn_ESR_arxiv.tex
\documentclass[twocolumn,notitlepage,prb,color,superscriptaddress,psfig,showpacs,amsmath,amssymb,nobibnotes,longbibliography]{revtex4-2}

\usepackage[colorlinks=true,citecolor=blue]{hyperref}%

\usepackage{epsf}
\usepackage{graphicx}
\usepackage{amssymb}
\usepackage{color}
\usepackage{float}
\usepackage{epstopdf}
\usepackage{natbib}
\usepackage{amsmath}
\usepackage{mathrsfs}

\usepackage{xspace}

\usepackage{soul}
\setstcolor{red}
\usepackage[version=4]{mhchem}

\usepackage[normalem]{ulem}

\definecolor{mygray}{cmyk}{0, 0, 0, 0.3}
\newcommand{\out}[1]{{\color{mygray}\sout{#1}}}

\def\BCPO{BiCoPO$_5$\xspace}

\begin{document}

	\title{Static magnetic and ESR spectroscopic properties of \\ the dimer-chain antiferromagnet BiCoPO$_5$}

\author{M.~Iakovleva}
\affiliation{Leibniz IFW Dresden, D-01069 Dresden, Germany}
\author{T.~Petersen}
\affiliation{Leibniz IFW Dresden, D-01069 Dresden, Germany}
\author{A.~Alfonsov}
\affiliation{Leibniz IFW Dresden, D-01069 Dresden, Germany}
\author{Y.~Skourski}	
\affiliation{Dresden High Magnetic Field Laboratory (HLD-EMFL), Helmholtz-Zentrum Dresden-Rossendorf, 01328 Dresden, Germany}
\author{H.-J.~Grafe}
\affiliation{Leibniz IFW Dresden, D-01069 Dresden, Germany}
\author{E.~Vavilova}
\affiliation{Zavoisky Physical-Technical Institute, FRC Kazan Scientific Center of RAS, 420029 Kazan, Russia}
\author{R.~Nath}
\affiliation{School of Physics, Indian Institute of Science Education and Research, Thiruvananthapuram-695551, India}
\author{L.~Hozoi}
\affiliation{Leibniz IFW Dresden, D-01069 Dresden, Germany}
\author{V.~Kataev}
\affiliation{Leibniz IFW Dresden, D-01069 Dresden, Germany}

	\date{\today}
	
	\begin{abstract}
		
We report a comprehensive study of the static susceptibility, high-field magnetization and high-frequency/high-magnetic field electron spin resonance (HF-ESR) spectroscopy of polycrystalline samples of the bismuth cobalt oxy-phosphate \BCPO. This compound features a peculiar spin system that can be considered as antiferromagnetic (AFM) chains built of pairs of ferromagnetically coupled Co spins and interconnected in all three spatial directions. It was previously shown that \BCPO orders antiferromagnetically at $T_{\rm N} \approx 10$\,K and this order can be continuously suppressed by magnetic field towards the critical value $\mu_0H_{\rm c} \approx 15$\,T. In our experiments we find
strongly enhanced magnetic moments and spectroscopic $g$ factors as compared to the expected spin-only values, suggesting a strong contribution of orbital magnetism for the Co$^{2+}$ ions.
This is quantitatively confirmed by {\it ab initio} quantum chemical calculations.
Within the AFM ordered phase, we observe a distinct field-induced magnetic phase transition. Its critical field rises to $\sim 6$\,T at $T \ll T_{\rm N}$. The HF-ESR spectra recorded at $T\ll T_{\rm N}$ are very rich comprising up to six resonance modes possibly of the multimagnonic nature that soften towards the critical region around 6\,T. Interestingly, we find that the Co moments are not yet
fully polarized at $H_{\rm c}$ which supports a theoretical proposal identifying $H_{\rm c}$ as the quantum critical point for the transition of the spin system in \BCPO to the quantum disordered state at stronger fields.

	\end{abstract}
	
	\maketitle
	
	\section{Introduction}

In the field of condensed matter quantum magnetism there is a recent rapidly growing interest in
Co-containing compounds, which
promise a rich material base for the realization of unconventional magnetic ground states. The expected interesting physics is related to a sufficiently strong spin-orbit coupling of the Co$^{2+}$ (3$d^7$) ion that entangles the spin $S = 3/2$ and the effective orbital momentum $l = 1$. In
crystalline lattices of certain symmetries this gives
rise to strongly anisotropic and frustrated exchange paths with resulting exotic spin orders or Kitaev spin liquid behaviors \cite{Liu2020,Kim2021}. The most studied materials in this respect are Na$_3$Co$_2$SbO$_6$ \cite{Viciu2007,Wong2016,Yan2019} and Na$_2$Co$_2$TeO$_6$ \cite{Viciu2007,Yao2020,Songvilay2020,Lin2021,Chen2021,Kim2021b,Samarakoon2021,Hong2021}, with honeycomb spin lattices, and ${\mathrm{Ba}}_{3}{\mathrm{CoSb}}_{2}{\mathrm{O}}_{9}$ \cite{Shirata2012,Susuki2013,Koutroulakis2015} and Na$_2$BaCo(PO$_4$)$_2$ \cite{Zhong19,Li2020,Lee2021,Wellm2021} with triangular spin lattice.

The magnetism of the title compound, the dimer-chain antiferromagnet \BCPO, has received so far lesser attention though, in general, materials containing spin dimer units interconnected in spin networks of different topology often demonstrate exotic quantum magnetic behaviors. 
%
%
To name a few examples, these are, e.g., a field-induced quantum phase transition into the transverse N{\'e}el ordered state in the double-chain compound TlCuCl$_3$ \cite{Tanaka2001}, or the Haldane spin-gap state and the field-induced Bose-Einstein condensation of magnons in the one-dimensional (1D) two-leg ladder spin-1/2 compounds IPA-CuCl$_3$ \cite{Masuda2006,Garlea2007} and CuBr$_4$(C$_5$H$_{12}$N)$_2$ \cite{Klanjsek2008,Thielemann2009} (for a comprehensive review see Ref.~\cite{Zapf2014}).

\begin{figure}[h]
	\centering
	\includegraphics[clip,width=0.65\columnwidth]{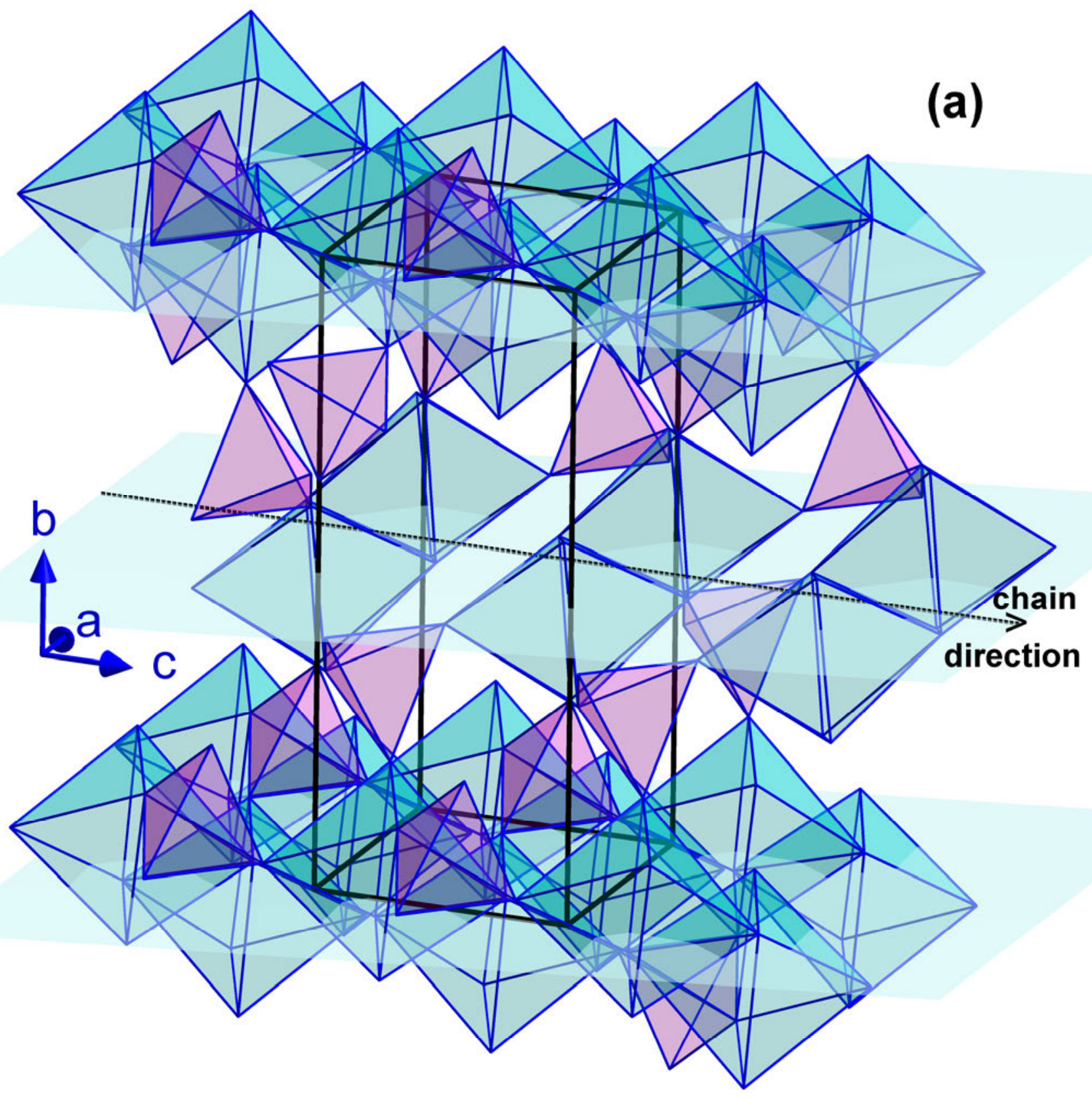}
	\includegraphics[clip,width=0.8\columnwidth]{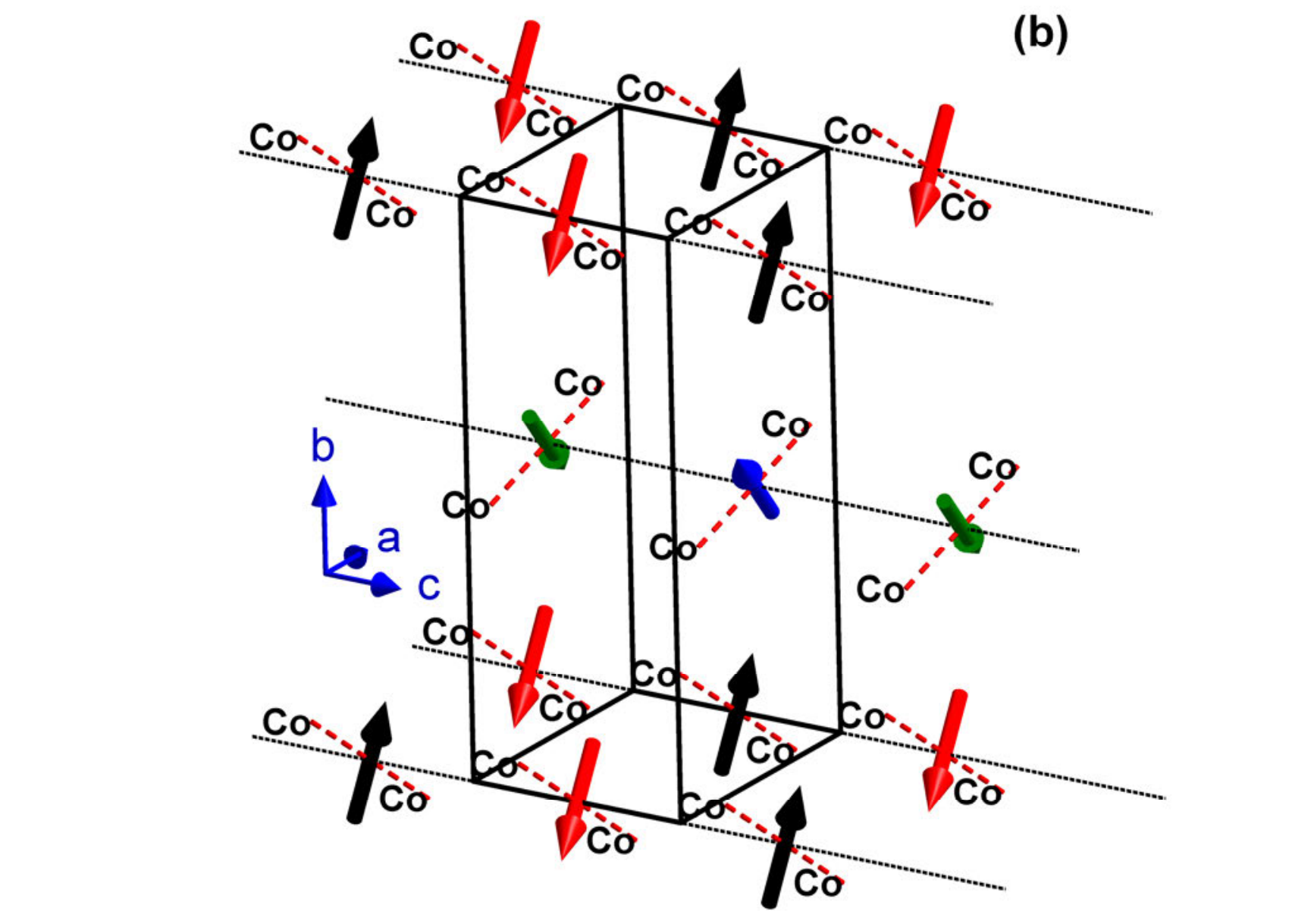}
	\caption{Crystal (a) and simplified magnetic (b) structures of \BCPO. CoO$_6$ octahedra and PO$_4$ tetrahedra are turquoise and pink colored, respectively. Bi atoms are not shown. Arrows depict the joint magnetic moment of the Co--Co dimers. (see Ref.~\cite{Mentre2008} and Sect.~\ref{ordered_state} for details)  }
	\label{fig:structure}
\end{figure}

The crystal crystal structure of \BCPO\ was extensively studied in Refs.~\cite{Nadir1999,Ketatni1999,Mentre2008} and is shown in Fig.~\ref{fig:structure}(a). The compound crystallizes in a monoclinic structure, space group $P2_{1}/n$\,(no. 14).
The main building units are CoO$_6$ octahedra sharing one edge and forming  structural dimers. The dimers are connected via PO$_4$ tetrahedra and form chains along the $c$~axis which are interconnected by PO$_4$ tetrahedra along the $a$ and $b$~axes. The Co spins in the structural dimer are exchange coupled into a ferromagnetic dimer and form antiferromagnetic chains along the $c$~axis that exhibit a noncollinear N\'eel order below $T_{\rm N} \approx 10$\,K due to weaker interchain couplings \cite{Mentre2008} [Fig.~\ref{fig:structure}(b)].

The basic magnetic properties of \BCPO were originally reported in Refs.~\cite{Nadir1999,Mentre2008} and re-investigated in Ref.~\cite{Mathews2013}. In the latter work also the magnetic specific heat $C_{\rm mag}(T)$ was measured in its dependence on the applied field. The $C_{\rm mag}(T,H)$ data evidenced a continuous reduction of $T_{\rm N}$ in an increasing magnetic field with the critical field for the suppression of the antiferromagnetic (AFM) order of $\mu_0H_{\rm c} \approx 15$\,T. Based on this finding Cai {\it et al.}~\cite{Cai2015} developed a theoretical model proposing \BCPO\ as a candidate material to show a quantum phase transition at $H_{\rm c}$ from the AFM ordered to a quantum disordered, spin-liquid like phase.

Exploration of the interesting physics that can be hosted by \BCPO\ requires first of all  thorough characterization of its magnetic properties in the $T - H$ parameter space. With this aim we carried out detailed measurements of the magnetic susceptibility in static fields up to 7\,T, of the magnetization in pulse magnetic fields up to 58\,T, and of high-frequency/high-field electron spin resonance (HF-ESR) spectroscopy in fields up to 16\,T on polycrystalline samples of the title compound.
These experiments were complemented by {\it ab initio} quantum chemical calculations for the Co-site multiplet structure, ground-state $g$~factors, and magnetization.
Our experimental and computational results evidence a strong entanglement of the spin and orbital momenta of the Co ions manifesting in large magnitudes of the effective moment, saturation magnetization and spectroscopic $g$~factor. We observed a rather sharp
 magnetic phase transition within the AFM phase of \BCPO\ at fields increasing up to $\sim 6$\,T at the lowest temperature. The low-energy magnetic excitations probed by HF-ESR reveal a rich spectrum of modes strongly dispersing in the magnetic field which indicate a possible multi-magnonic nature of these modes. We found that the full polarization of the Co moments is not achieved at the critical field $\mu_0H_{\rm c} \approx 15$\,T which supports the scenario of the phase transition to a partially spin-polarized quantum disordered state in \BCPO at $H > H_{\rm c}$.

\section{Experimental and \\ computational details}
\label{experimental}

Polycrystalline samples of \BCPO were prepared by a conventional solid state reaction technique described before in Ref.~\cite{Mathews2013}. Powder x-ray diffraction measurements confirmed the phase purity of the synthesized samples which were found to crystallize in a monoclinic structure with space group
$P2_{1}/n$ with the lattice parameters  $a \simeq 7.2441\,\AA$, $b \simeq 11.2828\,\AA$,
$c \simeq 5.2258\,\AA$, and $\beta \simeq 107.84^\circ$.

Static magnetic susceptibility $\chi(T,H)$ measurements were carried out with a Quantum Design MPMS-XL superconducting quantum interference device (SQUID) magnetometer in a temperature range $2-750$\,K and in magnetic fields up to 7\,T. The magnetization $M(H)$ curves were recorded in the Dresden High Magnetic Field Laboratory
with a pulsed magnet generating fields up to 60\,T. Magnetization of the sample was obtained by integration of the voltage induced in a compensated pick-up coil system surrounding the sample and calibrated by measuring the $M(H)$ curve of the sample up to 7\,T with the SQUID magnetometer (for further technical details see Ref.~\cite{Skourski2011}).

HF-ESR experiments in a frequency range $75-500$\,GHz were performed with a home-made multifrequency spectrometer which employed a vector network analyzer (PNA-X from Keysight Technologies) for measurements at frequencies up to 330\,GHz, and a modular Amplifier/Multiplier Chain  (AMC from Virginia Diodes, Inc.) in a combination with  a hot electron InSb bolometer (QMC Instruments) for measurements at higher frequencies. The sample was placed into a probe head operational in the transmission mode. The probe was mounted into a $^4$He variable temperature inset of a  superconducting magnet system (Oxford Instruments) generating fields up to 16\,T (see also Ref.~\cite{Alfonsov2021}).

\vspace{1em}

\begin{figure}[htbp]
	\centering
	\includegraphics[width=0.5\columnwidth]{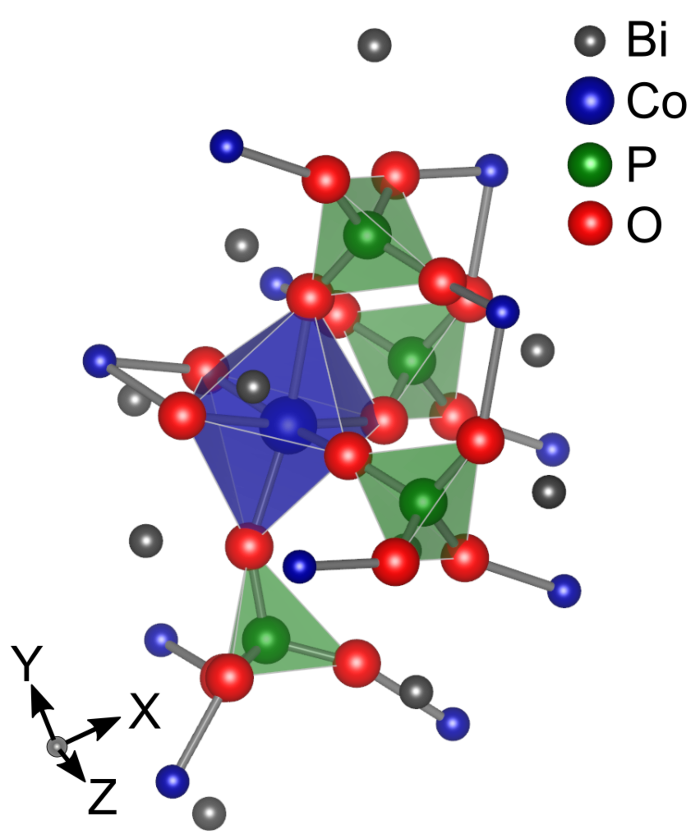}
	\caption{
\ce{[CoO6P4O12]^14-} cluster model
employed for the quantum chemical calculations of the Co-site multiplet structure.
The cluster atoms are indicated at the edges of the polyhedra.
As a boundary region between
the central cluster and the point charge surroundings
21 capped effective core potentials (cECPs) were used. (Bi: gray, Co: blue, P: green, O: red)
	}
	\label{fig:cluster}
\end{figure}

For insights into the Co 3$d$-level electronic structure, quantum chemical computational methods and an embedded-cluster material model were employed.
The quantum mechanical cluster itself, a \ce{[CoO6P4O12]^14-} fragment, is immersed within an array of point charges at lattice positions (see Fig.\:\ref{fig:cluster}).
With lattice parameters as derived from neutron powder diffraction by Mentre \textit{et al.} \cite{Mentre2008}, the point charge field  was optimized using the \textsc{Ewald} program \cite{Derenzo2000,Klintenberg2000}.
For the initial charge optimization, formal oxidation numbers of $+3$, $+2$, $+5$ and $-2$ were assumed for Bi, Co, P, and O, respectively.
A much more flexible description was adopted however for P and O species having to do with \ce{PO_4} tetrahedra directly linked to the central \ce{CoO6} octahedron --- those were explicitly included in the quantum mechanical cluster.
This ensures an accurate representation of the electronic charge distribution around the reference \ce{Co^2+} ion up to the third atomic coordination shell.
Between the quantum cluster and the point charges, a boundary region consisting of capped effective core potentials (cECPs) was devised, using the pseudopotentials of Küchle \cite{Kuechle1991} for \ce{Bi^3+} and of Dolg \cite{Dolg1987} for \ce{Co^2+}.
By introducing this additional region, artificial polarization of outer cluster orbitals toward adjacent positive point charges is prevented.
%
DKH-def2-TZVPP basis sets \cite{Weigend2005} were adopted for the central \ce{CoO6} part of the quantum mechanical cluster, while smaller DKH-def2-SVP basis sets \cite{Weigend2005} for the neighboring \ce{PO3}
fragments were used.
Excited state energies were estimated through \textit{ab initio} complete active space self-consistent field (CASSCF) theory \cite{Roos1987} by including the five $3d$ atomic orbitals of \ce{Co^2+} into the active space and averaging over all possible multiplets (10 quartets and 40 doublets).
Scalar relativistic properties were accounted for by enabling the second-order Douglas-Kroll-Hess (DKH) algorithm \cite{Douglas1974,Hess1986}.
Secondly, dynamical correlation was treated by $N$-electron valence second-order perturbation (NEVPT2) theory and multi-reference equation of motion (MR-EOM-T$\vert$T$^\dagger$-h-v) theory.
For NEVPT2, all internal orbitals were correlated, while for MR-EOM-T$\vert$T$^\dagger$-h-v localized internal orbitals comprising 18 2$p$ atomic orbitals of the nearest-neighbor oxygens and 3$s$ plus 3$p$ atomic orbitals of the Co ion were considered.
All calculations were performed with the program package Orca v5.0 \cite{Neese2022}.

\section{Experimental results}
\label{results}

\subsection{Susceptibility and magnetization}
\label{magnresults}

\begin{figure}[ht]
	\centering
	\includegraphics[clip,width=1.0\columnwidth]{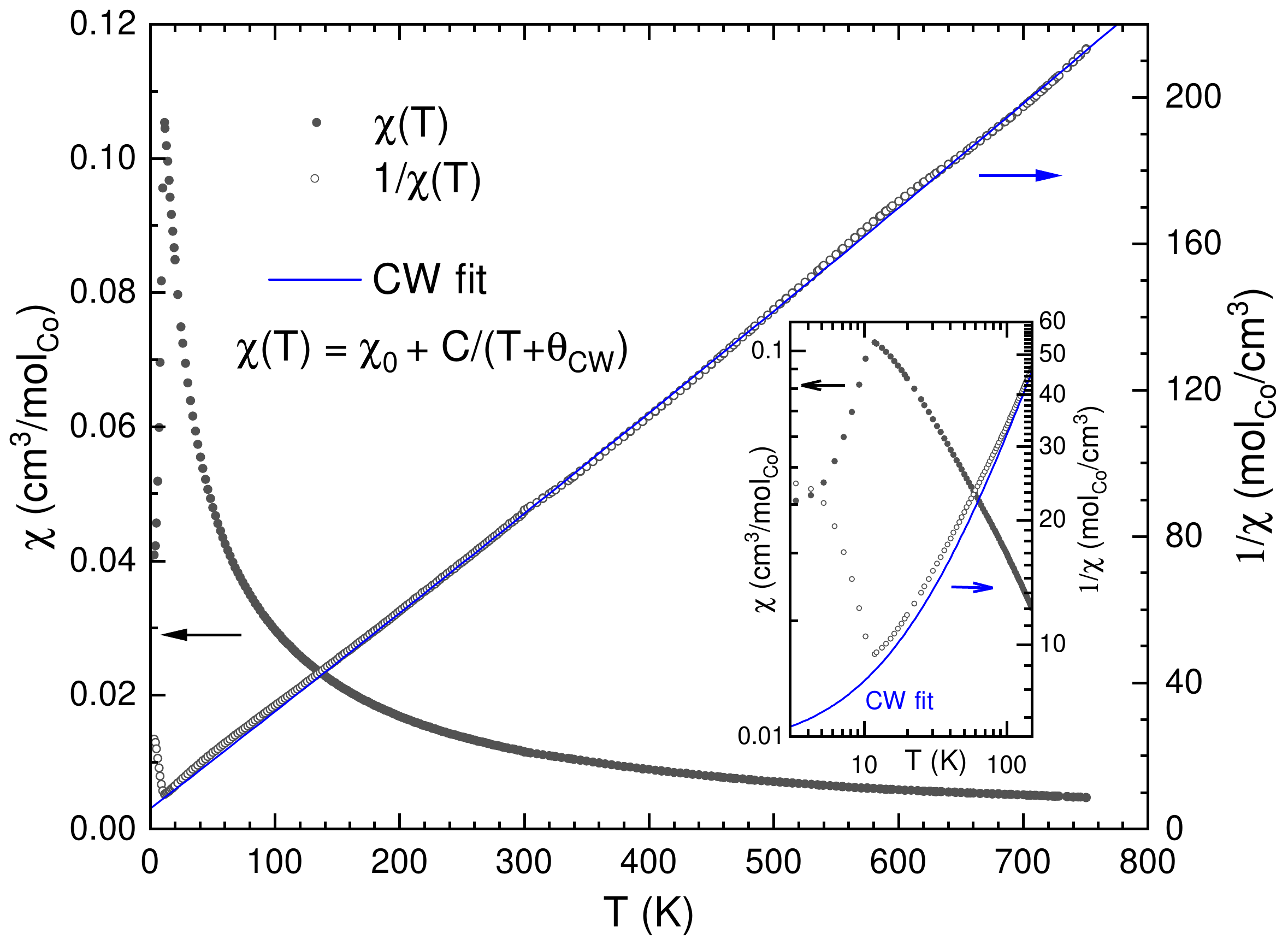}
	\caption{Temperature dependence of the static susceptibility $\chi(T)$ and its inverse $1/\chi(T)$ measured in a field of 0.1\,T (symbols). The solid line depicts the Curie-Weiss fit. The inset shows the expanded low temperature region on the log-log scale. }
	\label{fig:chi_T_1}
\end{figure}

The static susceptibility $\chi(T)$ of \BCPO as a function of temperature measured up to $T = 750$\,K at $\mu_0 H = 0.1$\,T is shown in Fig.~\ref{fig:chi_T_1}. The plot of $1/\chi(T)$ on the same Figure reveals a linear dependence down to $T\sim 100$\,K. It can be well fitted to the Curie-Weiss (CW) law $\chi(T) = \chi_0 + C/(T+T_{\rm CW}) $ yielding a small temperature independent contribution  $\chi_0 \simeq -2.14\cdot 10^-4$\,cm$^3$/mol$_{\rm Co}$, the Curie constant $C \simeq 3.79$\,cm$^3$K/mol$_{\rm Co}$ and the CW temperature $T_{\rm CW} \simeq 21$\,K. At lower temperatures the data deviate from the CW law [Fig.~\ref{fig:chi_T_1}(inset)] and the $\chi(T)$ dependence exhibits a peak at $T_{\rm m} = 11.8$\,K signaling a transition to the AFM ordered state. These results are in a reasonable agreement with the previous $\chi(T)$ measurements by Mathews {\it et al.}~\cite{Mathews2013}.
\begin{figure}[ht]
	\centering
	\includegraphics[clip,width=1.0\columnwidth]{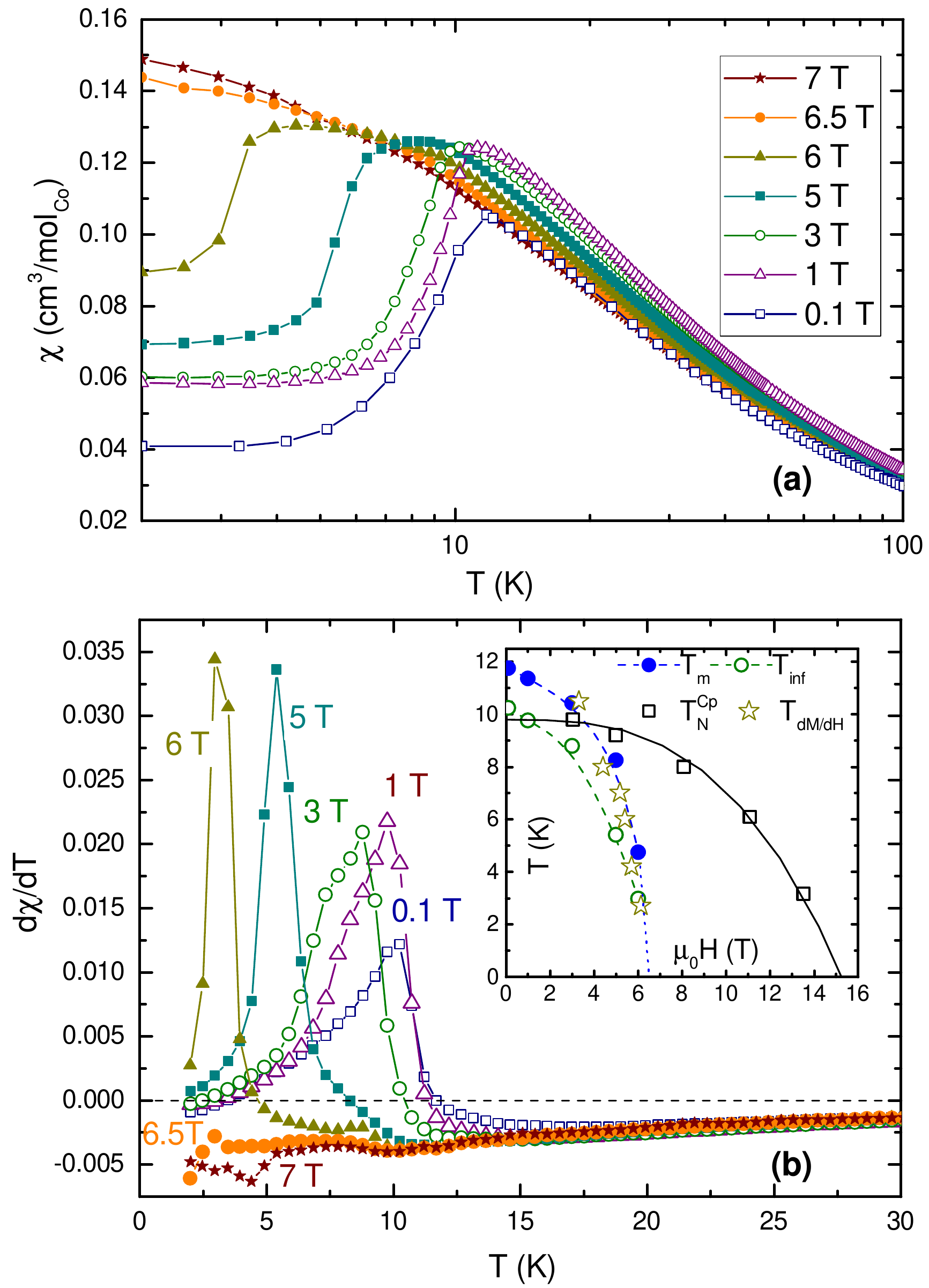}
	\caption{(a) Temperature dependence of the static susceptibility $\chi(T)$ at different applied fields (symbols connected with lines). (b) Derivatives of the respective $\chi(T)$ curves. The inset shows the field dependence of the temperature $T_{\rm m}$ of the maximum $\chi(T)$ corresponding to $d\chi/dT = 0$ (closed circles) and of the temperature $T_{\rm inf}$ of the inflection point of $\chi(T)$ corresponding to $d\chi/dT = max$ (open circles). Star symbols correspond to the position of the $dM/dH$ peak in Fig.~\ref{fig:M_H}.  Open squares are the ordering temperature $T_{\rm N}^{C{\rm p}}$ determined from the specific heat data in Ref.~\cite{Mathews2013}. The solid line depicts the fit to the phenomenological relation $T_{\rm N}(H) = T_{\rm N}(H = 0)[1-(H/H_{\rm c})^3]$ with $T_{\rm N}(H = 0) = 9.8$\,K and $\mu_0H_{\rm c}=15.3$\;T~\cite{Mathews2013}.}
	\label{fig:chi_T_2}
\end{figure}

\begin{figure}[ht]
	\centering
	\includegraphics[clip,width=1.0\columnwidth]{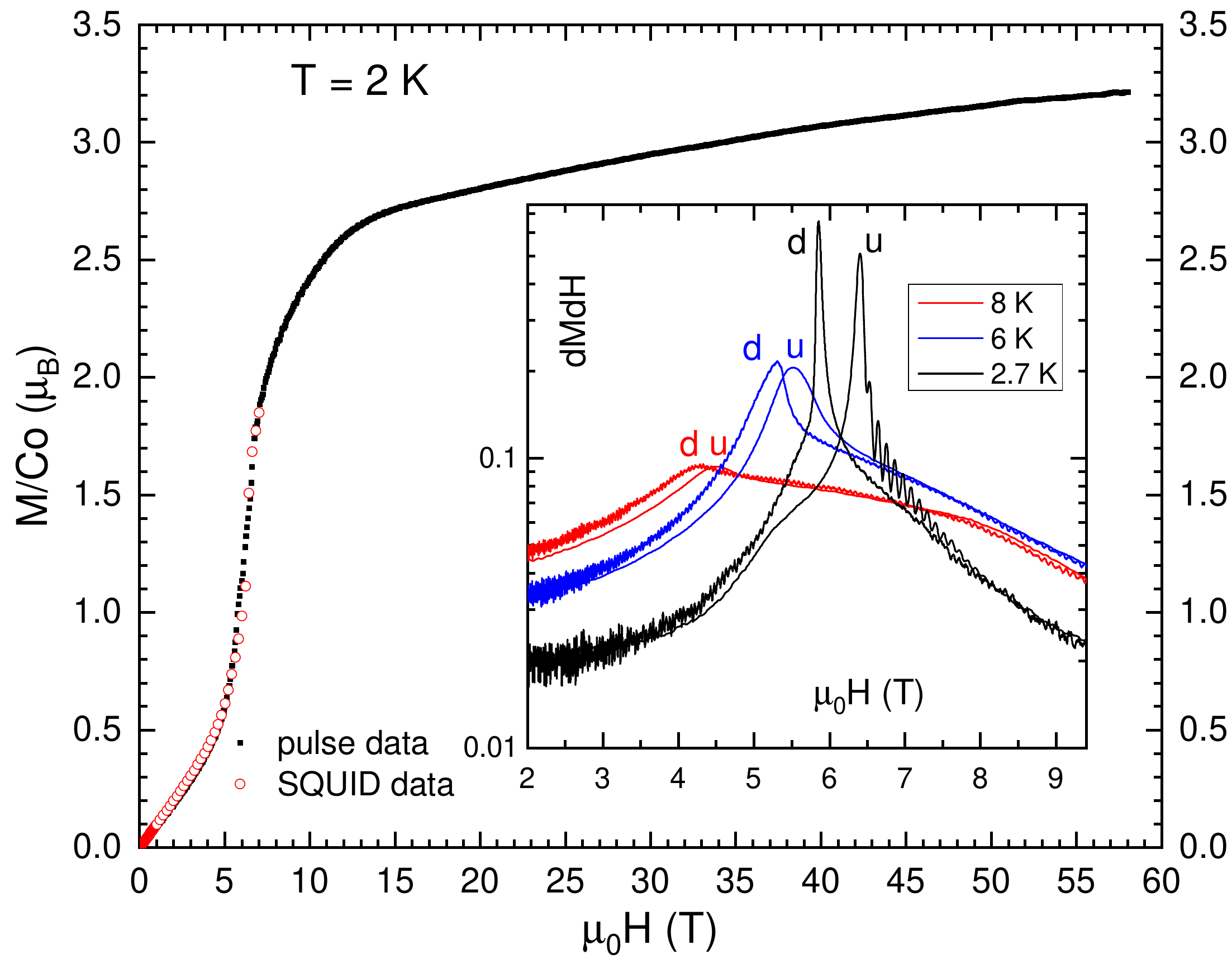}
	\caption{Field dependence of the magnetization $M(H)$ at $T=2$\,K measured in static fields up to 7\,T with SQUID (open circles) and in pulse fields up to 58\,T (filled squares). The latter curve is the average between the up- and down-field sweeps. The pulse data are scaled to the SQUID data. Inset: The $dM(H)/dH$ curves measured at selected temperatures. Labels "u" and "d" denote the up- and down-field sweep directions, respectively.}
	\label{fig:M_H}
\end{figure}

With increasing the field strength the peak of $\chi(T)$ broadens and shifts to lower temperatures and is not observed anymore above $\mu_0 H = 6$\,T [Fig.~\ref{fig:chi_T_2}(a)]. The peak temperature $T_{\rm m}$ can be accurately determined by plotting the derivative $d\chi(T)/dT$ where the maximum of $\chi(T)$ corresponds to $d\chi(T)/dT = 0$ and the inflection point $T_{\rm inf}$ of $\chi(T)$ at $T < T_{\rm m}$ corresponds to $d\chi(T)/dT = max$ [Fig.~\ref{fig:chi_T_2}(b)]. The results are shown in the inset to Fig.~\ref{fig:chi_T_2}(b) together with the field dependence of the critical N\'eel temperature of an AFM phase transition $T_{\rm N}^{C{\rm p}}(H)$ obtained from the specific heat measurements by Mathews {\it et al.}~\cite{Mathews2013}. There, the position of the $dM/dH$ peak at different temperatures shown in Fig.~\ref{fig:M_H} is presented as well. A comparison of these characteristic temperatures reveals that at small magnetic fields the inflection point of the $\chi(T)$ curve $T_{\rm inf}$ corresponds well with the critical N\'eel temperature $T_{\rm N}^{C{\rm p}}(H)$ whereas the peak  of the $\chi(T)$ dependence occurs at a slightly higher temperature $T_{\rm m}$.

%
%

%

The field dependence of the magnetization $M(H)$ of \BCPO at $T = 2$\,K
is shown in the main panel
of Fig.~\ref{fig:M_H}. After an initial shallow increase the $M(H)$ rises up above the inflection point $H_{\rm inf}\sim 6.3$\,T and gradually reaches the saturation value $M_{\rm sat} \simeq 3.2$\,$\mu_{\rm B}$/Co by approaching a field of 60\,T.
The differential magnetization  $dM(H)/dH$ at selected temperatures in the field range around the inflection point of the $M(H)$ curve  is shown in the inset. The position of the peak of $dM(H)/dH$ corresponds to $H_{\rm inf}$ at a given temperature. With decreasing the temperature the peak develops hysteresis. The average peak position between the up- and down-field sweeps is plotted in the inset to Fig.~\ref{fig:chi_T_2}.

\begin{figure}[ht]
	\centering
	\includegraphics[clip,width=0.6\columnwidth]{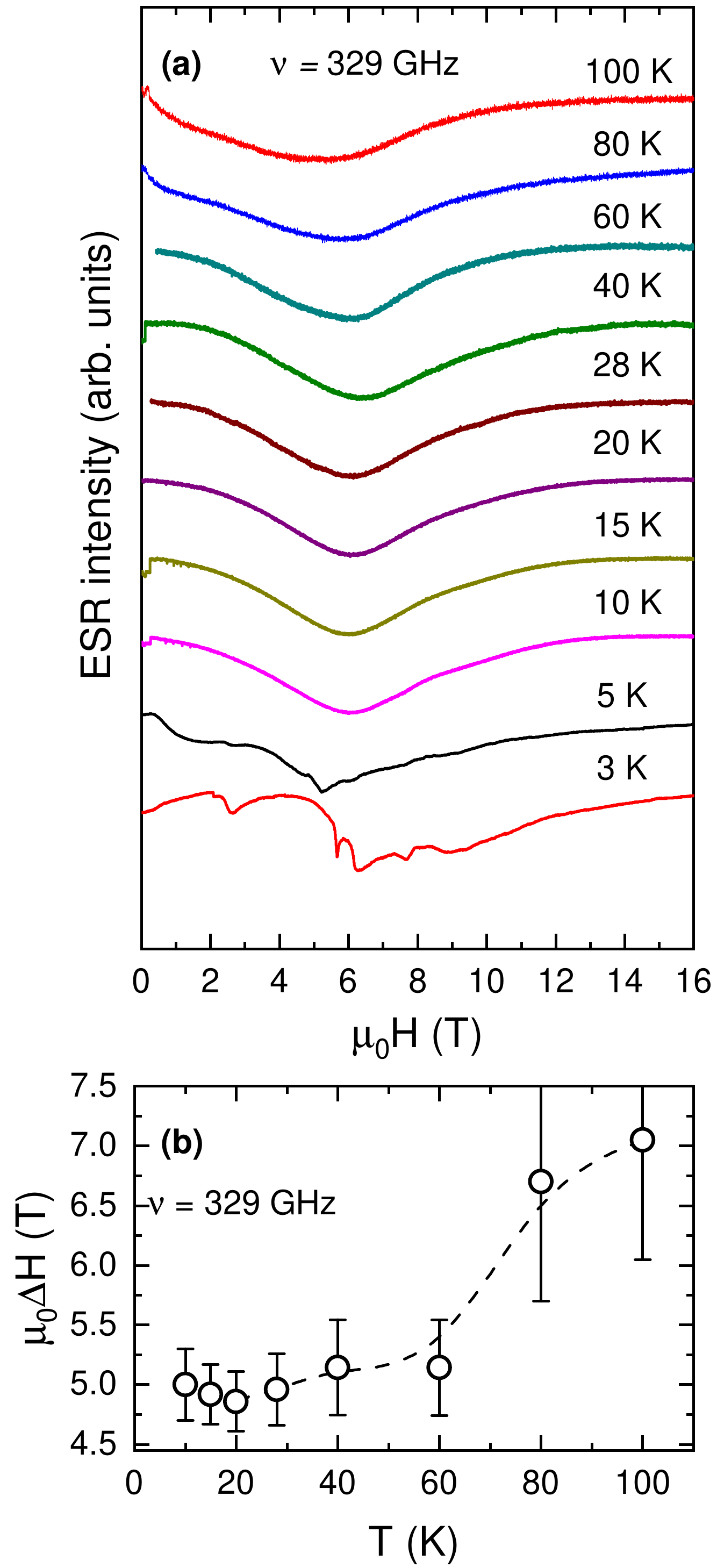}
	\caption{Temperature dependence of the ESR spectrum at $\nu =329$\,GHz (a) and of the width $\Delta H(T)$ at half-height of the ESR line in the paramagnetic state (b).  }
	\label{fig:ESR_T}
\end{figure}

\begin{figure}[ht]
	\centering
	\includegraphics[clip,width=0.9\columnwidth]{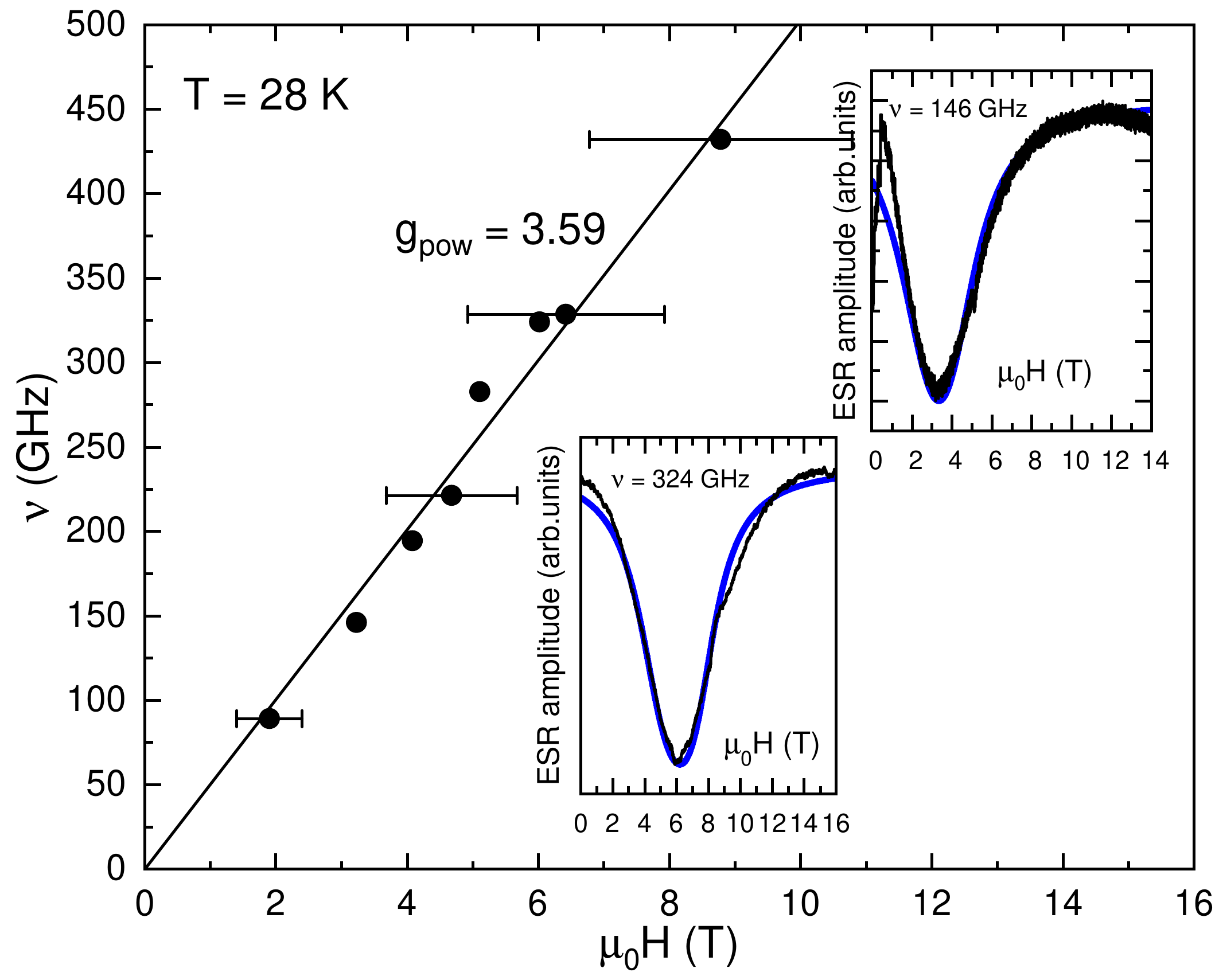}
	\caption{Position of the ESR peak at different frequencies at $T = 28$\,K. The solid line depicts the linear dependence $\nu = h^{-1}g_{\rm pow}\mu_0\mu_{\rm B}H$ with the powder-averaged $g$~factor $g_{\rm pow} = 3.59$. Insets show exemplary ESR signals (black) together with the powder-averaged model curves (blue). (see the text for details)   }
	\label{fig:ESR_freq}
\end{figure}

\subsection{ESR spectroscopy}
\label{ESRresults}
A very broad HF-ESR spectrum of \BCPO can be first observed at $T\sim 100$\,K. It narrows significantly by lowering the temperature $T \sim 60\,{\rm K} \gg T_{\rm N}$ possibly due to the depopulation of the excited multiplet states of the Co$^{2+}$ ion \cite{Wellm2021}.
Exemplary spectra recorded at the excitation frequency $\nu = 329$\,GHz are shown in Fig.~\ref{fig:ESR_T}(a) and the temperature dependence of their width $\Delta H(T)$ is plotted in Fig.~\ref{fig:ESR_T}(b). To obtain the spectroscopic $g$~factor in the paramagnetic state HF-ESR spectra at $T =28$\,K\,$ > T_{\rm N}$ and at various frequencies were collected. The relation between the position of the absorption peak and $\nu$ is plotted in Fig.~\ref{fig:ESR_freq}. The data can be fitted to the paramagnetic resonance condition $\nu = h^{-1}g_{\rm pow}\mu_0\mu_{\rm B}H$ yielding the powder-averaged $g$~factor $g_{\rm pow} \simeq 3.59$. Further, the HF-ESR signals were modeled with the powder-averaged Lorentzian line profiles with an anisotropic $g$~tensor. Assuming for simplicity the uniaxial $g$~factor anisotropy the modeling curves matched well with the experimental lineshapes at different frequencies with the $g$~tensor $g_\parallel = 4.94$ and $g_\perp = 2.91$ [Fig.~\ref{fig:ESR_freq}(inset)].
Its average, $\sqrt{(g_\parallel^2 +2g_\perp^2)/3} = 3.71$ is consistent with the $g_{\rm pow}$ value obtained from the fit in Fig.~\ref{fig:ESR_freq}.

\begin{figure}[ht]
	\centering
	\includegraphics[clip,width=0.9\columnwidth]{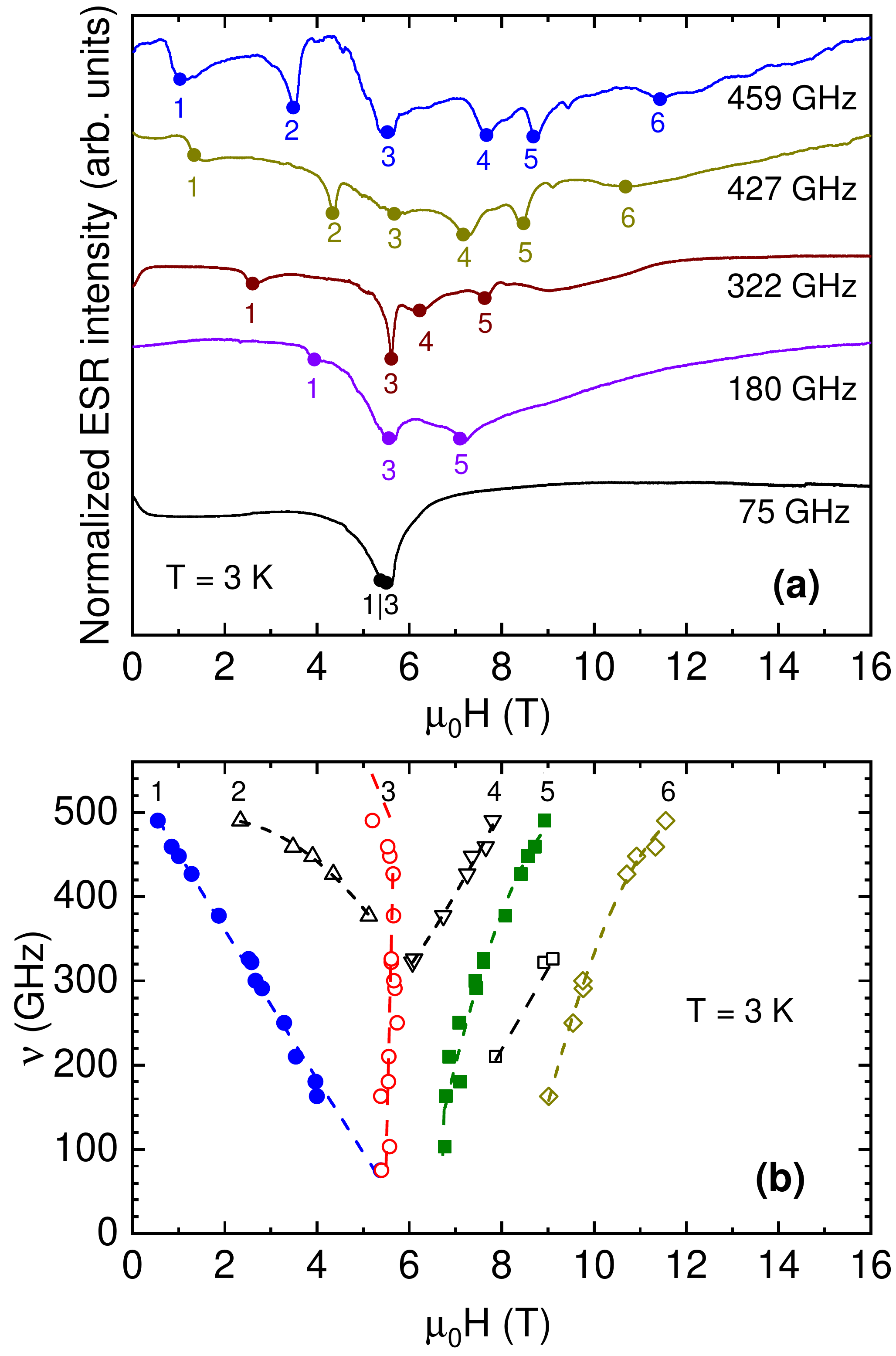}
	\caption{HF-ESR spectra at $T = 3$\,K and at different frequencies (a). Summary of the peak positions of the spectral lines which are grouped in branches labeled from 1 to 6 (b). Lines connecting the data points in (b) are the guides for the eye. }
	\label{fig:AFMR}
\end{figure}

By lowering the temperature below $T_{\rm N}$ the HF-ESR signal gradually transforms into a complex spectrum of multiple excitations. Its evolution at $T = 3$\,K with changing the frequency is shown in Fig.~\ref{fig:AFMR}(a). The spectrum converges to an almost single resonance line at the minimum frequency $\nu = 75$\,GHz. The resonance fields of the observed excitations are summarized in the $\nu$ {\it versus} $H$ diagram in Fig.~\ref{fig:AFMR}(b). They can be classified into six resonance branches as indicated there. It appears that the critical field region around 6\,T identified in the analysis of the magnetization data [Fig.~\ref{fig:chi_T_2}(inset)] also plays a special role for the HF-ESR excitations in the AFM ordered state of \BCPO.

\section{Discussion}
\label{discussion}

\subsection{Paramagnetic state}
\label{paramgnetic_state}

\subsubsection{Experimental phenomenology}

In the phenomenological description of the static susceptibility $\chi(T)$ of \BCPO with the Curie-Weiss law the Curie constant $C$ is defined as $C = N_{\rm Co}N_{\rm A}M_{\rm eff}^2/3k_{\rm B}$.
Here, $N_{\rm Co}$, $N_{\rm A}$ and $k_{\rm B}$ are the number of Co ions in the formula unit, the Avogadro number and the Boltzmann constant, respectively. $M_{\rm eff}$ is the effective magnetic moment $M_{\rm eff} = g[j(j+1)]^{1/2}\mu_{\rm B}$ in the units of Bohr magneton and $g$ and $j$ are the $g$~factor and the effective spin. 

With $C = 3.79$\,cm$^3$K/mol$_{\rm Co}$ from the CW fit (Fig.~\ref{fig:chi_T_1}) one obtains $M_{\rm eff}=5.51\,\mu_{\rm B}$/Co. One the other hand, $M_{\rm eff}$ is related to the saturation magnetization $M_{\rm sat} = gj\mu_{\rm B}$ as $M_{\rm eff}^2 = M_{\rm sat}^2 + gM_{\rm sat}$. According to this relation with $M_{\rm sat} = 3.2$\,$\mu_{\rm B}$/Co obtained from the $M(H)$ dependence (Fig.~\ref{fig:M_H}) the $g$~factor should amount to $g = 6.29$ in a strong disagreement with the experimental value $g_{\rm pow} = 3.59$ (Fig.~\ref{fig:ESR_freq}) and with the predictions of the canonical ligand theory \cite{Ballhausen1962}.
In its framework the $g$~factor in the ground state of an isolated Co$^{2+}$ ion in a regular octahedral coordination  should be isotropic and amount to $g_{\rm cub} = 10/3 + k$ where $k$ is the orbital reduction factor $k < 1$ due to the covalency effect. For the ionic bond $k$ equals to 1. The $g$ factor much larger than the spin-only value $g = 2$ results from the combined action of a strong li\-gand cubic field and the spin-orbit coupling.
Distortion of  the ligand octahedron causes anisotropy of the $g$~factor while the mean $g$~factor remains close to $g_{\rm cub}$ \cite{AbragamBleaney,Pilbrow1990}. In the case of \BCPO the $g_{\rm pow}$ is smaller than the pure ionic value of 4.33 which may indicate a substantial covalency of the Co--O bonds. Similar reduced mean $g$~factors were found, e.g., for some covalently bonded Co(II) inorganic complexes \cite{Pilbrow1990,Marts2017}.

A consistency between the static magnetic data and the $g$~factor obtained in the HF-ESR measurements can be restored assuming that the ferromagnetic coupling between the spins in the Co$_2$O$_{10}$ structural dimers in the AFM ordered spin structure of \BCPO\ (see Sect.~\ref{ordered_state}) is so strong that the ferromagnetic (FM) dimers are present also in the paramagnetic state. In this scenario, deviation of the experimental 1/$\chi(T)$ dependence from the CW law at $T > 500$\,K (Fig.~\ref{fig:chi_T_1}) might indicate the breakdown of the FM dimers.
Recalculating the Curie constant per mole of the Co--Co FM dimers $C_{\rm Co_2} = 2C = 7.58$\,cm$^3$K/mol$_{\rm Co_2}$ yields the effective moment of the dimer $M_{\rm eff}^{\rm Co_2}=7.79\,\mu_{\rm B}$/Co$_2$. Considering the above relation between $M_{\rm eff}$ and the saturation magnetization $M_{\rm sat}$, with $g_{\rm pow} = 3.59$ one obtains $M_{\rm sat}^{\rm Co_2}=6.2\,\mu_{\rm B}$/Co$_2$ which corresponds to 3.1$\mu_{\rm B}$ per one Co site, in a close agreement with the experiment.

%
%

\subsubsection{{\it Ab initio} multiplet structure}

The multiplet structure of the \ce{[CoO6P4O12]^14-} cluster for the minimal CAS(7e,5o) space (i.e., all possible configurations related to distributing seven electrons within the five Co 3$d$ orbitals) is illustrated in Table \ref{tab:coo6_cas75}.
Due to the strong distortion of the \ce{CoO6} octahedron, the ground state $^4T_\text{1g}$ term \cite{Sugano1970} is split into three non-degenerate states.
The magnitude of these splittings is significant, up to 90 meV.
Even richer structure is obtained after including spin-orbit couplings (SOCs) in the computations.
The emerging picture is that of six Kramers doublets (KDs) grouped into two sets of states --- two KDs defining a low-energy scale of 25 meV for the on-site excitations and another four KDs implying excitation energies of 90--170 meV.
Similar $t_\text{2g}^5e_\text{g}^2$ multiplet structures were found in \ce{Co^2+} complexes of interest in the field of single molecule magnets \cite{Lloret2008,Murrie2010,Atanasov2015,Kumar2020}.

Correcting the CASSCF approach by NEVPT2 and MR-EOM-T$\vert$T$^\dagger$-h-v schemes yields only minor corrections within the group of states related to the $t_\text{2g}^5e_\text{g}^2$ electron configuration.
The corrections are however important for higher-lying multiplets.
In particular, the $t_\text{2g}^5e_\text{g}^2$ states are stabilized with respect to the $t_\text{2g}^4e_\text{g}^3$ states (correlation effects involving O-to-Co charge transfer are more important for the former since there is less charge in the $\sigma$-type Co $e_\text{g}$ orbitals).
There are additionally ``differential'' effects within the group of $t_\text{2g}^4e_\text{g}^3$ states, e.g., modification of the sequence of the $^4A_\text{2g}$ and $^2E_\text{g}$ terms.
\begin{table*}[htbp]
	\caption{Co$^{2+}$ 3$d^7$ multiplet structure as computed for a \ce{[CoO6P4O12]^14-} cluster (CAS(7e,5o)/DKH-def2-TZVPP).
Notations corresponding to octahedral $O_\text{h}$ point-group symmetry are used, although the actual symmetry is much lower.
	}
	\begin{ruledtabular}
		\begin{tabular}{l c c c c}
			$d^7$ term & CASSCF [meV] & CASSCF+SOC [meV] & NEVPT2+SOC [meV] & MR-EOMT$\vert$T$^\dagger$-h-v+SOC [meV] \\
			\colrule
			$^4T_\text{1g}$ &    0 &    0,   25 &    0,   24 &    0,   25 \\
			&   63 &   86,  128 &   93,  132 &   95,  135 \\
			&   91 &  147,  160 &  158,  170 &  160,  173 \\
			$^4T_\text{2g}$ &  606 &  652,  662 &  811,  820 &  758,  768 \\
			&  657 &  701,  708 &  893,  899 &  825,  833 \\
			&  739 &  784,  799 &  989, 1002 &  911,  925 \\
			$^4A_\text{2g}$ & 1389 & 1444, 1445 & 1819, 1820 & 1675, 1676 \\
			$^2E$           & 1909 & 1952 &       1421       & 1716 \\
			& 2149 & 2190 &       1757       & 2035 \\
			$^2T_\text{1g}$ & 2414 & 2451 &       2168       & 2083 \\
			...             & ... &  ... & ... & ... \\
		\end{tabular}
	\end{ruledtabular}
	\label{tab:coo6_cas75}
\end{table*}

\begin{figure}[htbp]
	\centering
	\includegraphics[width=0.9\columnwidth]{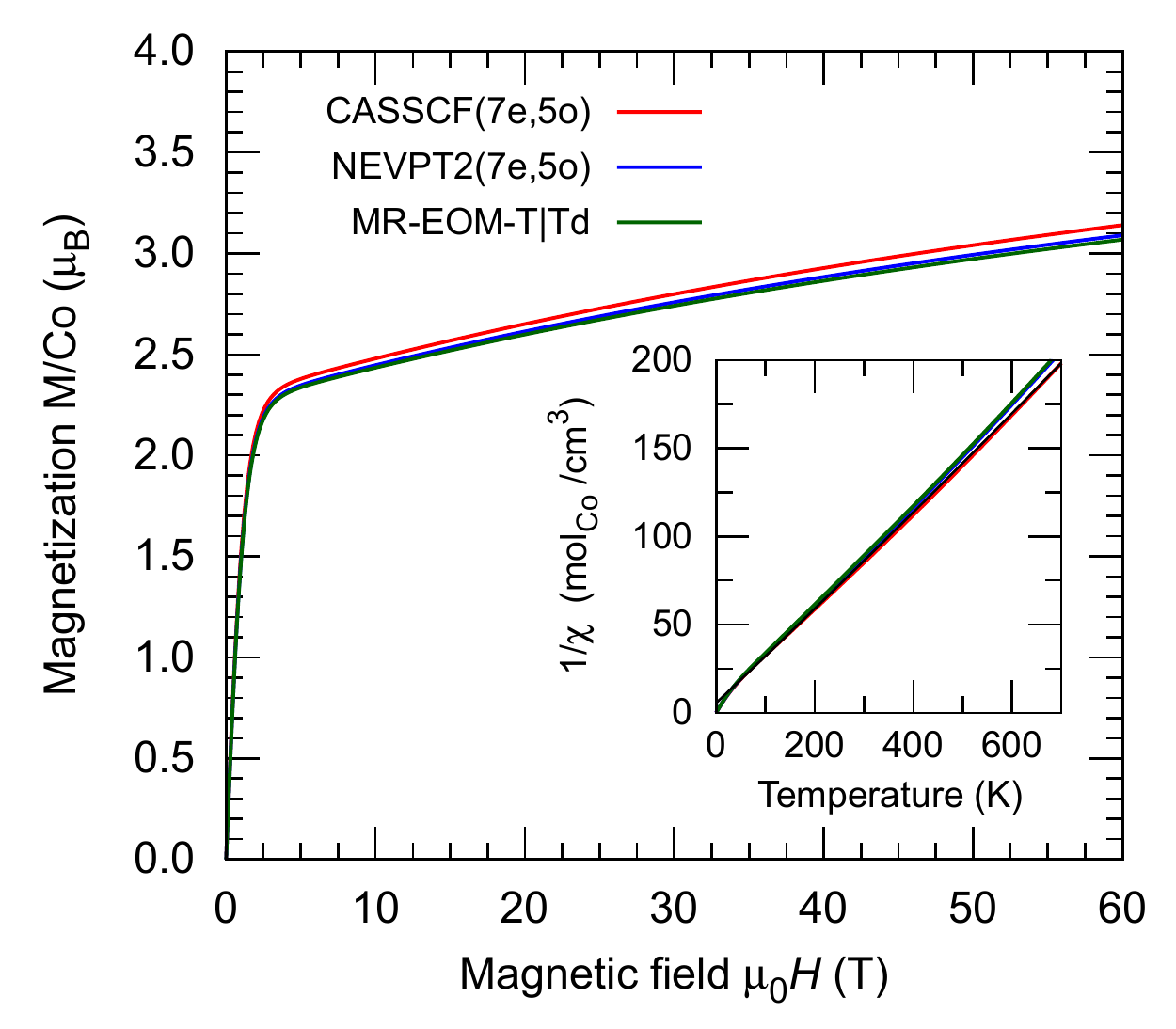}
	\caption{
Field dependence of the magnetization $M$ calculated at $T=2\:\text{K}$ for the \ce{[CoO6P4O12]^14-} cluster.
		Inset: Computed temperature dependence of the inverse magnetic susceptibility $1/\chi(T)$ in an applied field of 0.1\:T (CASSCF(7e,5o): red, NEVPT2: blue, MR-EOM-T$\vert$T$^\dagger$-h-v: green, experimental CW fit: black).
	}
	\label{fig:1magn_2susc}
\end{figure}

For direct connection with experiment, using the tools available with the Orca package \cite{Neese2022} and on the basis of the multiplet spectrum summarized in Table\:\ref{tab:coo6_cas75}, we computed magnetization and magnetic susceptibility curves (see Fig.\:\ref{fig:1magn_2susc}).
Focusing first on the magnetization curve, no considerable difference is observed between the different applied computational methods.
Compared to the experimental $M(H)$ dependence shown in Fig.~\ref{fig:M_H}, the magnetization curve calculated for $T = 2$\,K initially goes steeper for low applied fields, since inter-octahedral magnetic interactions yielding long-range
\out{a} 
AFM order
\out{in \BCPO} 
(Sect.~\ref{ordered_state}) are not captured with a single-octahedron cluster model.
However, the experimentally determined saturation value $M_\text{sat}^\text{exp}$=$3.2\:\mu_\text{B}/\text{Co}$ corresponding to
\out{a} 
practically full alignment of the Co moments by approaching a field of 60\,T
is reasonably well reproduced in theory
($M_\text{sat}^\text{theor}$=$3.09\:\mu_\text{B}$ for NEVPT2).
The calculated inverse magnetic susceptibility $1/\chi$ is shown in the inset of Fig.\:\ref{fig:1magn_2susc}.
The experimental and computational data coincide over a wide temperature range from about 100 to 400 K.

Subsequently, the
average $g_\text{avg}$ factor was determined by mapping the SO eigenvalues onto
a $\tilde{S}$=$1/2$ effective Hamiltonian \cite{Chibotaru2012}.
A value of $g_\text{avg}$=$4.30$ was found by NEVPT2 ($g_\text{avg}$=$4.37$ by CASSCF(7e,5o)).
The NEVPT2 and CASSCF values are quite similar, in agreement with the magnetization and magnetic susceptibility curves shown in Fig.\:\ref{fig:1magn_2susc}.
However, the $g_\text{avg}$ factor deviates somewhat from the
value determined by HF-ESR.
Test CASSCF computations on a larger, two-octahedra cluster \ce{[Co2O10P8O24]^24-} confirm the localized nature of the Co valence orbitals found in the single-octahedron calculations
and the presence of ferromagnetic exchange between the two edge-sharing octahedra.
More involved post-CASSCF computations for quantitative analysis of the inter-site magnetic couplings are left for future work.

\subsection{Magnetically ordered state}
\label{ordered_state}

\begin{figure}[ht]
	\centering
	\includegraphics[clip,width=0.9\columnwidth]{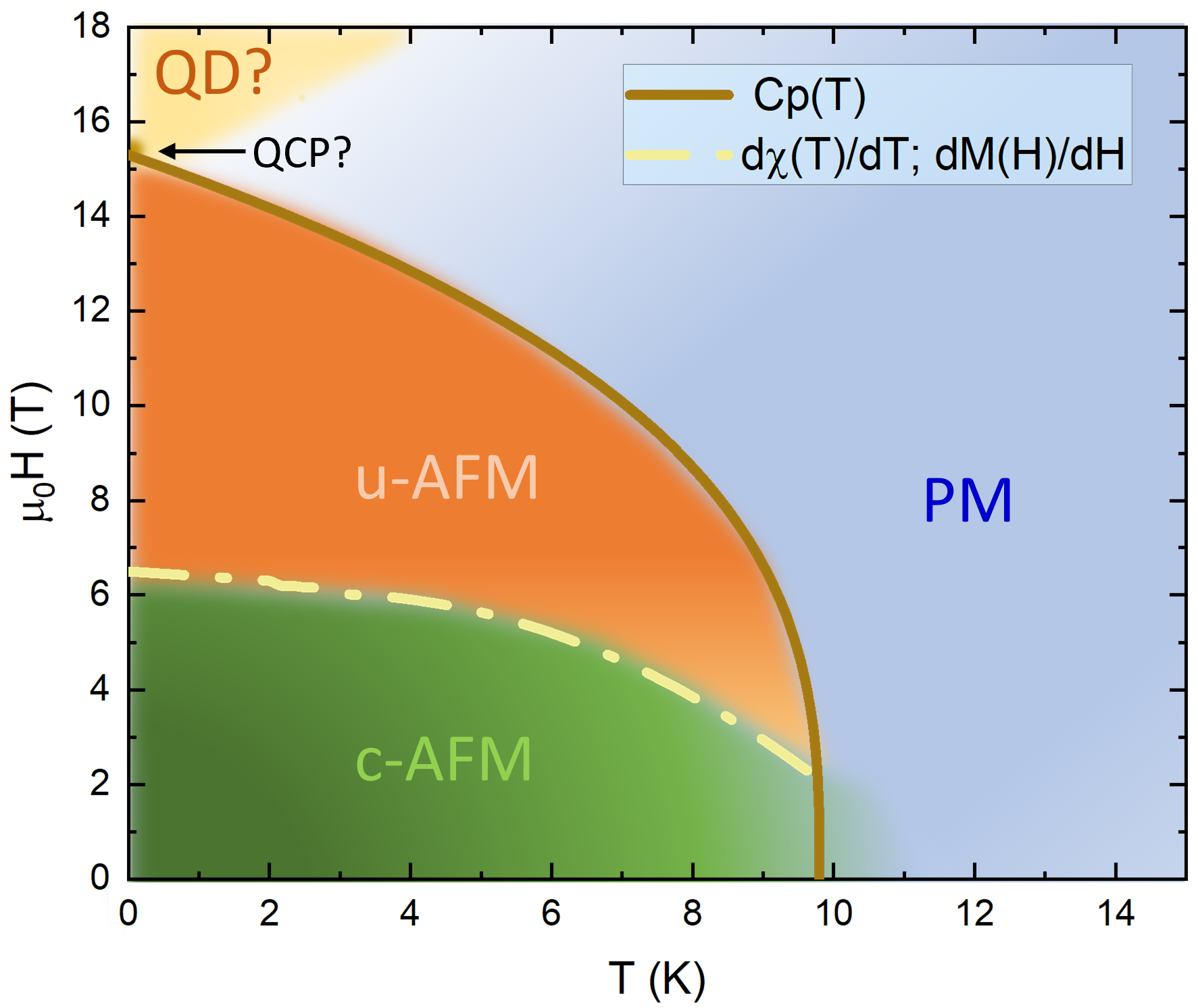}
	\caption{Schematic $H-T$ phase diagram of \BCPO. The solid line delimits the paramagnetic (PM) and AFM ordered phases according to the specific heat data in Ref.~\cite{Mathews2013} [Fig.~\ref{fig:chi_T_2}(inset)]. The dash-dotted line corresponds to the spin reorientation transition according to the static susceptibility $d\chi(T)/dT$ and magnetization $dM(H)/dT$ data in Figs.~\ref{fig:chi_T_2} and \ref{fig:M_H} and delimits the weakly magnetized, compensated AFM phase (c-AFM) at low fields and the stronger magnetized, uncompensated AFM phase (u-AFM) at higher fields. The quantum disordered phase (QD) anticipated in Ref.~\cite{Cai2015} emerges above the quantum critical point (QCP) at $\mu_0H > 15$\,T.      }
	\label{fig:phasediagram}
\end{figure}

The magnetic structure of \BCPO\ was solved by powder neutron diffraction experiments by Mentre {\it et al.}~\cite{Mentre2008}. In this work also the topological analysis of the relative strength and sign of different Co--Co exchange paths was performed. The strongest FM coupling was found in the Co$_2$O$_{10}$ structural dimer which thus can be considered as a  magnetic unit with the magnetic moment with the size twice of that of the Co$^{2+}$ ion. The dimer moments build up AFM chains along the $c$~axis which are interconnected by weaker interchain exchange interactions [Fig.~\ref{fig:structure}(b)].

The dimer moments in the neighboring $ac$~planes of the chain are tilted by an angle of 47.6$^\circ$ (132.4$^\circ$) with respect to each other. These moments can be grouped in four sublattices [black, red, green and blue in Fig.~\ref{fig:structure}(b)] which together form a four-sublattice noncollinear and nonorthogonal AFM ordered structure referred to as a compensated AFM ordered phase (c-AFM) with zero net magnetization in the limit $(H,T) \rightarrow 0$ (Fig.~\ref{fig:phasediagram}). The ordered moment of the Co$^{2+}$ ion was found to be 3.5\,$\mu_{\rm B}$ in a reasonable agreement with the saturation magnetization in 58\,T where the almost full polarization of the Co moments in \BCPO\ can be achieved (Fig.~\ref{fig:M_H}).

Application of the external magnetic field tilts the sublattices which causes a shift of the peak in $\chi(T)$ at $T_{\rm m}$ and of the inflection point $T_{\rm inf}$ below the peak to lower temperatures and broadens the peak (Fig.~\ref{fig:chi_T_2}). The sharp peak of the differential magnetization at $T_{\rm dM/dH}$ [Fig.~\ref{fig:M_H}(inset)] shows a similar $T$\,{\it versus}\,$H$ behavior as $T_{\rm m}(H)$ and $T_{\rm inf}(H)$ which altogether may indicate a field induced magnetic phase transition in the spin system of \BCPO. This is a transition from the c-AFM state to a stronger magnetized uncompensated AFM state (u-AFM) within the AFM ordered phase below the $T_{\rm N}(H)$ phase line determined in the specific heat measurements in Ref.~\cite{Mathews2013} (Fig.~\ref{fig:phasediagram}).
Usually a strong field induced increase of the magnetization $M(H)$ is ascribed to the spin-flop transition if ${\bf H}$ is applied parallel to the magnetic easy~axis (direction of spins) of a collinear antiferromagnet. The hysteresis character of this transition [Fig.~\ref{fig:M_H}(inset)] suggests that it is of first order, which is typical for a spin-flop transition \cite{Bogdanov2007}.
However, in the noncollinear magnetic structure of \BCPO\ there are at least two easy axes and the sublattices cannot be flopped at a unique critical field value even in a single crystal. Powder averaging should further broaden the transition region as indeed observed for the dimer-chain antiferromagnet BiMnVO$_5$ possessing the crystal structure  similar to \BCPO\ \cite{Chowki2016}. The sharpness of the $dM/dH$ peak at $T = 2.7$\,K is thus somewhat surprising and may indicate a more complex nature of the field induced magnetic transition in \BCPO.

We notice here an interesting similarity of the $H-T$ phase diagram of \BCPO\ sketched in Fig.~\ref{fig:phasediagram} with that of the Kitaev-like honeycomb magnet Na$_2$Co$_2$TeO$_6$ \cite{Yao2020}. There, a sharp first-order field-induced spin reorientation transition to a stronger magnetized state  was observed at $\mu_0 H \sim 6$\,T  which evolves into a tentative quantum paramagnetic state at still higher fields, qualitatively similar to the scenario proposed for \BCPO in Ref.~\cite{Cai2015} (see Fig.~\ref{fig:phasediagram}). 

The complexity of the ground state of \BCPO\ manifests also in the HF-ESR excitation spectrum at $T = 3\,{\rm K} \ll T_{\rm N}$ comprising multiple lines that disperse differently in the magnetic field (Fig.~\ref{fig:AFMR}). Line 3 in the HF-ESR spectrum in Fig.~\ref{fig:AFMR}(a) and the position of the corresponding branch 3 in Fig.~\ref{fig:AFMR}(b) do not depend on frequency. The occurrence of line 3 is most likely due to a nonresonant change of the microwave absorption by the sample due to the change of its magnetization. Indeed, its position coincides with the peak of the differential magnetization [Fig.~\ref{fig:M_H}(inset)] and thus branch 3 delimits the excitations in the two magnetic phases of \BCPO\ at this low temperature. Branches 1--2 on the low field side and branches 4--6 on the high-field side of the $\nu(H)$ diagram soften towards the critical field region around $\sim 6$\,T. 
It should be noted that in the more conventional three-dimensional (3D) antiferromagnet BaCo$_2$(SeO$_3$)$_3\cdot$3H$_2$O containing structural Co-dimer chains only two ESR modes softening towards the critical field of the spin-flop transition were observed, that correspond to the expected single-magnon excitations of a two-sublattice collinear antiferromagnet \cite{Liu2022}.
%
%

It would be tempting to attribute the steep resonance branches in Fig.~\ref{fig:AFMR} to complex multiparticle excitations in \BCPO. For example, branch 1 could be well approximated by the softening two-magnon excitation $\nu_1 = \Delta_1(H=0) - h^{-1}g_{\rm pow}\mu_0\mu_{\rm B}H\Delta S_{\rm z}$ with the double spin flip $\Delta S_{\rm z} =2$
and the zero-field excitation gap $\Delta_1(H=0) = 565$\,GHz. The multiple modes in \BCPO\ bear some similarities with the ESR excitation spectrum of $\alpha$-RuCl$_3$, a prominent Heisenberg-Kitaev spin liquid candidate, where the resonance lines strongly dispersing in the magnetic field and softening towards the critical field for the transition from an AFM ordered phase to a partially polarized disordered phase  were attributed to multimagnonic  excitations \cite{Wang2017,Ponomaryov2017,Wellm2018,Ponomaryov2020,Sahasrabudhe2020}. Such excitations were reproduced in numerical calculations based on the Heisenberg-Kitaev Hamiltonian with sizable  off-diagonal anisotropic exchange $\Gamma$ terms \cite{Winter2018}. 
It has been recently discussed that the signatures of the Kitaev physics could be found  not only  in anisotropic  quasi-two-dimensional  honeycomb and triangular spin lattices but also, e.g., in the 1D spin chains \cite{Yang2020,Luo2021,Macedo2022}, or in the 3D cobalt spinels and pyrochlores \cite{Motome2020}. Given a significant spin-orbit entanglement found for Co$^{2+}$ ions in \BCPO\ such H--K--$\Gamma$ Hamiltonian might possibly be relevant for this material as well. However, the exact assignment of the ESR branches would require complex calculations of the magnon modes in a four-sublattice noncollinear antiferromagnet which is beyond the scope of the present work.

Finally, it is not possible to draw a definite conclusion on the existence of the theoretically predicted new resonance mode emerging out of the tentative quantum critical point $\mu_0 H_{\rm c} = 15$\,T where the bulk AFM order in \BCPO\ is suppressed \cite{Cai2015}. Considering that the low temperature magnetization is still far from being saturated at $H_{\rm c}$ (Fig.~\ref{fig:M_H}), one can conjecture that this field might indeed be a special critical point for the transition of the spin system in \BCPO to a partially polarized, disordered state (Fig.~\ref{fig:phasediagram}). According to Cai {\it al.}~\cite{Cai2015} the predicted excitation mode should have zero energy (zero frequency) at $H_{\rm c}$ and acquire a linear in field energy gap at $H > H_{\rm c}$. Unfortunately with the current experimental HF-ESR setup limited to fields $\mu_0H \leq 16$\,T and frequencies $\nu \geq 75$\,GHz this mode might lie outside the available  $\nu - H$ observation window and remain undetected.

\section{Conclusions}

In summary, we discussed the results of detailed measurements of the static susceptibility, magnetization and high-frequency ESR in the paramagnetic and AFM ordered state of the dimer-chain antiferromagnet \BCPO.
The Co$^{2+}$ 3$d^7$ multiplet structure was calculated with many-body, {\it ab initio} quantum chemical methods.
The experimentally observed large magnetic moment and a large and anisotropic spectroscopic $g$~factor of the  Co$^{2+}$ ion are consistently reproduced in the calculations and explained as being a consequence of
the combined action of sizable spin-orbit coupling and strong low-symmetry ligand fields.
In the phenomenological analysis of the static magnetic data indications of the existence of
ferromagnetic Co-dimers in the paramagnetic state up to high temperatures were found.
However, the quantum chemical computations do not provide evidence for the formation of quasi-molecular orbitals, as found in various 4$d$ and 5$d$ compounds \cite{Mazin2012,Komleva2020,Tsirlin2022}. This should be related with the more compact nature of 3$d$ orbitals and with the presence of stronger correlations in 3$d$ systems \cite{Komleva2020}.

Low temperature magnetization and HF-ESR data give evidence for a rather sharp field-induced magnetic phase transition within the AFM ordered phase of \BCPO taking place at a field of $\sim 6$\,T at the lowest temperature. HF-ESR spectra at $T = 3$\,K\,$\ll T_{\rm N}$ feature multiple resonance modes tentatively ascribed to multi-magnonic excitations that soften towards this critical 6\,T field region. The magnetization is not yet saturated at a critical field for the suppression of the AFM order $\mu_0 H_{\rm c} = 15$\,T \cite{Mathews2013} favoring a theoretical proposal in Ref.~\cite{Cai2015} on the occurrence of a quantum disordered state in \BCPO above $\mu_0 H_{\rm c}$ (Fig.~\ref{fig:phasediagram}). However, the new resonance mode emergent in this theory at $\mu_0 H_{\rm c}$ was not detected, possibly due to the technical limitations of the used HF-ESR setup.

The properties of the field-induced u-AFM phase at $6-7< \mu_0 H < 15$\,T (Fig.~\ref{fig:phasediagram}) are not captured in the theory in Ref.~\cite{Cai2015} and require further clarification. At present it is not clear  whether this phase is long-range or only short-range ordered, possibly due to the enhanced quantum fluctuations in proximity to the conjectured quantum critical point at $\mu_0 H_{\rm c} \sim 15$\,T, and whether the c-AFM$\rightarrow$u-AFM transition is of a classical or quantum nature.
Altogether our results call for further exploration of the interesting magnetic phase diagram of \BCPO with local spin probes and development of theoretical models of exchange interactions and magnetic excitations which could pinpoint the envisaged field-induced quantum criticality in \BCPO.

\section*{Acknowledgments}

\noindent
We thank S.~Aswartham for useful discussions, and U.~Nitzsche and S.~Ga\ss\ for technical assistance. We acknowledge the support of the HLD at HZDR, member of the European Magnetic Field Laboratory (EMFL).
T.\;P.~and L.\;H.~acknowledge financial support from the Deutsche Forschungsgemeinschaft (DFG) (Projects No. 437124857 and 468093414).


\input{BiCoPO5_magn_ESR_arxiv.bbl}

\end{document}

%% file: BiCoPO5_magn_ESR_arxiv.bbl
%

%% file: BiCoPO5_magn_ESR_arxiv.bbl
\begin{thebibliography}{67}%
\makeatletter
\providecommand \@ifxundefined [1]{%
 \@ifx{#1\undefined}
}%
\providecommand \@ifnum [1]{%
 \ifnum #1\expandafter \@firstoftwo
 \else \expandafter \@secondoftwo
 \fi
}%
\providecommand \@ifx [1]{%
 \ifx #1\expandafter \@firstoftwo
 \else \expandafter \@secondoftwo
 \fi
}%
\providecommand \natexlab [1]{#1}%
\providecommand \enquote  [1]{``#1''}%
\providecommand \bibnamefont  [1]{#1}%
\providecommand \bibfnamefont [1]{#1}%
\providecommand \citenamefont [1]{#1}%
\providecommand \href@noop [0]{\@secondoftwo}%
\providecommand \href [0]{\begingroup \@sanitize@url \@href}%
\providecommand \@href[1]{\@@startlink{#1}\@@href}%
\providecommand \@@href[1]{\endgroup#1\@@endlink}%
\providecommand \@sanitize@url [0]{\catcode `\\12\catcode `\$12\catcode
  `\&12\catcode `\#12\catcode `\^12\catcode `\_12\catcode `\%12\relax}%
\providecommand \@@startlink[1]{}%
\providecommand \@@endlink[0]{}%
\providecommand \url  [0]{\begingroup\@sanitize@url \@url }%
\providecommand \@url [1]{\endgroup\@href {#1}{\urlprefix }}%
\providecommand \urlprefix  [0]{URL }%
\providecommand \Eprint [0]{\href }%
\providecommand \doibase [0]{https://doi.org/}%
\providecommand \selectlanguage [0]{\@gobble}%
\providecommand \bibinfo  [0]{\@secondoftwo}%
\providecommand \bibfield  [0]{\@secondoftwo}%
\providecommand \translation [1]{[#1]}%
\providecommand \BibitemOpen [0]{}%
\providecommand \bibitemStop [0]{}%
\providecommand \bibitemNoStop [0]{.\EOS\space}%
\providecommand \EOS [0]{\spacefactor3000\relax}%
\providecommand \BibitemShut  [1]{\csname bibitem#1\endcsname}%
\let\auto@bib@innerbib\@empty
\bibitem [{\citenamefont {Liu}\ \emph {et~al.}(2020)\citenamefont {Liu},
  \citenamefont {Chaloupka},\ and\ \citenamefont {Khaliullin}}]{Liu2020}%
  \BibitemOpen
  \bibfield  {author} {\bibinfo {author} {\bibfnamefont {H.}~\bibnamefont
  {Liu}}, \bibinfo {author} {\bibfnamefont {J.}~\bibnamefont {Chaloupka}},\
  and\ \bibinfo {author} {\bibfnamefont {G.}~\bibnamefont {Khaliullin}},\
  }\bibfield  {title} {\bibinfo {title} {{Kitaev Spin Liquid in $3d$ Transition
  Metal Compounds}},\ }\href {https://doi.org/10.1103/PhysRevLett.125.047201}
  {\bibfield  {journal} {\bibinfo  {journal} {Phys. Rev. Lett.}\ }\textbf
  {\bibinfo {volume} {125}},\ \bibinfo {pages} {047201} (\bibinfo {year}
  {2020})}\BibitemShut {NoStop}%
\bibitem [{\citenamefont {Kim}\ \emph {et~al.}(2021{\natexlab{a}})\citenamefont
  {Kim}, \citenamefont {Kim},\ and\ \citenamefont {Park}}]{Kim2021}%
  \BibitemOpen
  \bibfield  {author} {\bibinfo {author} {\bibfnamefont {C.}~\bibnamefont
  {Kim}}, \bibinfo {author} {\bibfnamefont {H.-S.}\ \bibnamefont {Kim}},\ and\
  \bibinfo {author} {\bibfnamefont {J.-G.}\ \bibnamefont {Park}},\ }\bibfield
  {title} {\bibinfo {title} {{Spin-orbital entangled state and realization of
  Kitaev physics in 3$d$ cobalt compounds: a progress report}},\ }\href
  {https://doi.org/10.1088/1361-648x/ac2d5d} {\bibfield  {journal} {\bibinfo
  {journal} {J. Condens. Matter Phys.}\ }\textbf {\bibinfo {volume} {34}},\
  \bibinfo {pages} {023001} (\bibinfo {year} {2021}{\natexlab{a}})}\BibitemShut
  {NoStop}%
\bibitem [{\citenamefont {Viciu}\ \emph {et~al.}(2007)\citenamefont {Viciu},
  \citenamefont {Huang}, \citenamefont {Morosan}, \citenamefont {Zandbergen},
  \citenamefont {Greenbaum}, \citenamefont {McQueen},\ and\ \citenamefont
  {Cava}}]{Viciu2007}%
  \BibitemOpen
  \bibfield  {author} {\bibinfo {author} {\bibfnamefont {L.}~\bibnamefont
  {Viciu}}, \bibinfo {author} {\bibfnamefont {Q.}~\bibnamefont {Huang}},
  \bibinfo {author} {\bibfnamefont {E.}~\bibnamefont {Morosan}}, \bibinfo
  {author} {\bibfnamefont {H.}~\bibnamefont {Zandbergen}}, \bibinfo {author}
  {\bibfnamefont {N.}~\bibnamefont {Greenbaum}}, \bibinfo {author}
  {\bibfnamefont {T.}~\bibnamefont {McQueen}},\ and\ \bibinfo {author}
  {\bibfnamefont {R.}~\bibnamefont {Cava}},\ }\bibfield  {title} {\bibinfo
  {title} {{Structure and basic magnetic properties of the honeycomb lattice
  compounds Na$_2$Co$_2$TeO$_6$ and Na$_3$Co$_2$SbO$_6$}},\ }\href
  {https://doi.org/https://doi.org/10.1016/j.jssc.2007.01.002} {\bibfield
  {journal} {\bibinfo  {journal} {J. Solid State Chem.}\ }\textbf {\bibinfo
  {volume} {180}},\ \bibinfo {pages} {1060} (\bibinfo {year}
  {2007})}\BibitemShut {NoStop}%
\bibitem [{\citenamefont {Wong}\ \emph {et~al.}(2016)\citenamefont {Wong},
  \citenamefont {Avdeev},\ and\ \citenamefont {Ling}}]{Wong2016}%
  \BibitemOpen
  \bibfield  {author} {\bibinfo {author} {\bibfnamefont {C.}~\bibnamefont
  {Wong}}, \bibinfo {author} {\bibfnamefont {M.}~\bibnamefont {Avdeev}},\ and\
  \bibinfo {author} {\bibfnamefont {C.~D.}\ \bibnamefont {Ling}},\ }\bibfield
  {title} {\bibinfo {title} {{Zig-zag magnetic ordering in honeycomb-layered
  Na$_3$Co$_2$SbO$_6$}},\ }\href
  {https://doi.org/https://doi.org/10.1016/j.jssc.2016.07.032} {\bibfield
  {journal} {\bibinfo  {journal} {J. Solid State Chem.}\ }\textbf {\bibinfo
  {volume} {243}},\ \bibinfo {pages} {18} (\bibinfo {year} {2016})}\BibitemShut
  {NoStop}%
\bibitem [{\citenamefont {Yan}\ \emph {et~al.}(2019)\citenamefont {Yan},
  \citenamefont {Okamoto}, \citenamefont {Wu}, \citenamefont {Zheng},
  \citenamefont {Zhou}, \citenamefont {Cao},\ and\ \citenamefont
  {McGuire}}]{Yan2019}%
  \BibitemOpen
  \bibfield  {author} {\bibinfo {author} {\bibfnamefont {J.-Q.}\ \bibnamefont
  {Yan}}, \bibinfo {author} {\bibfnamefont {S.}~\bibnamefont {Okamoto}},
  \bibinfo {author} {\bibfnamefont {Y.}~\bibnamefont {Wu}}, \bibinfo {author}
  {\bibfnamefont {Q.}~\bibnamefont {Zheng}}, \bibinfo {author} {\bibfnamefont
  {H.~D.}\ \bibnamefont {Zhou}}, \bibinfo {author} {\bibfnamefont {H.~B.}\
  \bibnamefont {Cao}},\ and\ \bibinfo {author} {\bibfnamefont {M.~A.}\
  \bibnamefont {McGuire}},\ }\bibfield  {title} {\bibinfo {title} {{Magnetic
  order in single crystals of
  ${\mathrm{Na}}_{3}{\mathrm{Co}}_{2}{\mathrm{SbO}}_{6}$ with a honeycomb
  arrangement of
  ${3\mathrm{d}}^{7}\phantom{\rule{0.28em}{0ex}}{\mathrm{Co}}^{2+}$ ions}},\
  }\href {https://doi.org/10.1103/PhysRevMaterials.3.074405} {\bibfield
  {journal} {\bibinfo  {journal} {Phys. Rev. Materials}\ }\textbf {\bibinfo
  {volume} {3}},\ \bibinfo {pages} {074405} (\bibinfo {year}
  {2019})}\BibitemShut {NoStop}%
\bibitem [{\citenamefont {Yao}\ and\ \citenamefont {Li}(2020)}]{Yao2020}%
  \BibitemOpen
  \bibfield  {author} {\bibinfo {author} {\bibfnamefont {W.}~\bibnamefont
  {Yao}}\ and\ \bibinfo {author} {\bibfnamefont {Y.}~\bibnamefont {Li}},\
  }\bibfield  {title} {\bibinfo {title} {{Ferrimagnetism and anisotropic phase
  tunability by magnetic fields in
  ${\mathrm{Na}}_{2}{\mathrm{Co}}_{2}{\mathrm{TeO}}_{6}$}},\ }\href
  {https://doi.org/10.1103/PhysRevB.101.085120} {\bibfield  {journal} {\bibinfo
   {journal} {Phys. Rev. B}\ }\textbf {\bibinfo {volume} {101}},\ \bibinfo
  {pages} {085120} (\bibinfo {year} {2020})}\BibitemShut {NoStop}%
\bibitem [{\citenamefont {Songvilay}\ \emph {et~al.}(2020)\citenamefont
  {Songvilay}, \citenamefont {Robert}, \citenamefont {Petit}, \citenamefont
  {Rodriguez-Rivera}, \citenamefont {Ratcliff}, \citenamefont {Damay},
  \citenamefont {Bal\'edent}, \citenamefont {Jim\'enez-Ruiz}, \citenamefont
  {Lejay}, \citenamefont {Pachoud}, \citenamefont {Hadj-Azzem}, \citenamefont
  {Simonet},\ and\ \citenamefont {Stock}}]{Songvilay2020}%
  \BibitemOpen
  \bibfield  {author} {\bibinfo {author} {\bibfnamefont {M.}~\bibnamefont
  {Songvilay}}, \bibinfo {author} {\bibfnamefont {J.}~\bibnamefont {Robert}},
  \bibinfo {author} {\bibfnamefont {S.}~\bibnamefont {Petit}}, \bibinfo
  {author} {\bibfnamefont {J.~A.}\ \bibnamefont {Rodriguez-Rivera}}, \bibinfo
  {author} {\bibfnamefont {W.~D.}\ \bibnamefont {Ratcliff}}, \bibinfo {author}
  {\bibfnamefont {F.}~\bibnamefont {Damay}}, \bibinfo {author} {\bibfnamefont
  {V.}~\bibnamefont {Bal\'edent}}, \bibinfo {author} {\bibfnamefont
  {M.}~\bibnamefont {Jim\'enez-Ruiz}}, \bibinfo {author} {\bibfnamefont
  {P.}~\bibnamefont {Lejay}}, \bibinfo {author} {\bibfnamefont
  {E.}~\bibnamefont {Pachoud}}, \bibinfo {author} {\bibfnamefont
  {A.}~\bibnamefont {Hadj-Azzem}}, \bibinfo {author} {\bibfnamefont
  {V.}~\bibnamefont {Simonet}},\ and\ \bibinfo {author} {\bibfnamefont
  {C.}~\bibnamefont {Stock}},\ }\bibfield  {title} {\bibinfo {title} {{Kitaev
  interactions in the Co honeycomb antiferromagnets
  ${\mathrm{Na}}_{3}{\mathrm{Co}}_{2}{\mathrm{SbO}}_{6}$ and
  ${\mathrm{Na}}_{2}{\mathrm{Co}}_{2}{\mathrm{TeO}}_{6}$}},\ }\href
  {https://doi.org/10.1103/PhysRevB.102.224429} {\bibfield  {journal} {\bibinfo
   {journal} {Phys. Rev. B}\ }\textbf {\bibinfo {volume} {102}},\ \bibinfo
  {pages} {224429} (\bibinfo {year} {2020})}\BibitemShut {NoStop}%
\bibitem [{\citenamefont {Lin}\ \emph {et~al.}(2021)\citenamefont {Lin},
  \citenamefont {Jeong}, \citenamefont {Kim}, \citenamefont {Wang},
  \citenamefont {Huang}, \citenamefont {Masuda}, \citenamefont {Asai},
  \citenamefont {Itoh}, \citenamefont {Guenther}, \citenamefont {Russina},
  \citenamefont {Lu}, \citenamefont {Sheng}, \citenamefont {Wang},
  \citenamefont {Wang}, \citenamefont {Wang}, \citenamefont {Ren},
  \citenamefont {Xi}, \citenamefont {Tong}, \citenamefont {Ling}, \citenamefont
  {Liu}, \citenamefont {Wu}, \citenamefont {Mei}, \citenamefont {Qu},
  \citenamefont {Zhou}, \citenamefont {Wang}, \citenamefont {Park},
  \citenamefont {Wan},\ and\ \citenamefont {Ma}}]{Lin2021}%
  \BibitemOpen
  \bibfield  {author} {\bibinfo {author} {\bibfnamefont {G.}~\bibnamefont
  {Lin}}, \bibinfo {author} {\bibfnamefont {J.}~\bibnamefont {Jeong}}, \bibinfo
  {author} {\bibfnamefont {C.}~\bibnamefont {Kim}}, \bibinfo {author}
  {\bibfnamefont {Y.}~\bibnamefont {Wang}}, \bibinfo {author} {\bibfnamefont
  {Q.}~\bibnamefont {Huang}}, \bibinfo {author} {\bibfnamefont
  {T.}~\bibnamefont {Masuda}}, \bibinfo {author} {\bibfnamefont
  {S.}~\bibnamefont {Asai}}, \bibinfo {author} {\bibfnamefont {S.}~\bibnamefont
  {Itoh}}, \bibinfo {author} {\bibfnamefont {G.}~\bibnamefont {Guenther}},
  \bibinfo {author} {\bibfnamefont {M.}~\bibnamefont {Russina}}, \bibinfo
  {author} {\bibfnamefont {Z.}~\bibnamefont {Lu}}, \bibinfo {author}
  {\bibfnamefont {J.}~\bibnamefont {Sheng}}, \bibinfo {author} {\bibfnamefont
  {L.}~\bibnamefont {Wang}}, \bibinfo {author} {\bibfnamefont {J.}~\bibnamefont
  {Wang}}, \bibinfo {author} {\bibfnamefont {G.}~\bibnamefont {Wang}}, \bibinfo
  {author} {\bibfnamefont {Q.}~\bibnamefont {Ren}}, \bibinfo {author}
  {\bibfnamefont {C.}~\bibnamefont {Xi}}, \bibinfo {author} {\bibfnamefont
  {W.}~\bibnamefont {Tong}}, \bibinfo {author} {\bibfnamefont {L.}~\bibnamefont
  {Ling}}, \bibinfo {author} {\bibfnamefont {Z.}~\bibnamefont {Liu}}, \bibinfo
  {author} {\bibfnamefont {L.}~\bibnamefont {Wu}}, \bibinfo {author}
  {\bibfnamefont {J.}~\bibnamefont {Mei}}, \bibinfo {author} {\bibfnamefont
  {Z.}~\bibnamefont {Qu}}, \bibinfo {author} {\bibfnamefont {H.}~\bibnamefont
  {Zhou}}, \bibinfo {author} {\bibfnamefont {X.}~\bibnamefont {Wang}}, \bibinfo
  {author} {\bibfnamefont {J.-G.}\ \bibnamefont {Park}}, \bibinfo {author}
  {\bibfnamefont {Y.}~\bibnamefont {Wan}},\ and\ \bibinfo {author}
  {\bibfnamefont {J.}~\bibnamefont {Ma}},\ }\bibfield  {title} {\bibinfo
  {title} {{Field-induced quantum spin disordered state in spin-1/2 honeycomb
  magnet Na$_2$Co$_2$TeO$_6$}},\ }\href
  {https://doi.org/10.1038/s41467-021-25567-7} {\bibfield  {journal} {\bibinfo
  {journal} {Nat. Commun.}\ }\textbf {\bibinfo {volume} {12}},\ \bibinfo
  {pages} {5559} (\bibinfo {year} {2021})}\BibitemShut {NoStop}%
\bibitem [{\citenamefont {Chen}\ \emph {et~al.}(2021)\citenamefont {Chen},
  \citenamefont {Li}, \citenamefont {Hu}, \citenamefont {Hu}, \citenamefont
  {Yue}, \citenamefont {Sutarto}, \citenamefont {He}, \citenamefont {Iida},
  \citenamefont {Kamazawa}, \citenamefont {Yu}, \citenamefont {Lin},\ and\
  \citenamefont {Li}}]{Chen2021}%
  \BibitemOpen
  \bibfield  {author} {\bibinfo {author} {\bibfnamefont {W.}~\bibnamefont
  {Chen}}, \bibinfo {author} {\bibfnamefont {X.}~\bibnamefont {Li}}, \bibinfo
  {author} {\bibfnamefont {Z.}~\bibnamefont {Hu}}, \bibinfo {author}
  {\bibfnamefont {Z.}~\bibnamefont {Hu}}, \bibinfo {author} {\bibfnamefont
  {L.}~\bibnamefont {Yue}}, \bibinfo {author} {\bibfnamefont {R.}~\bibnamefont
  {Sutarto}}, \bibinfo {author} {\bibfnamefont {F.}~\bibnamefont {He}},
  \bibinfo {author} {\bibfnamefont {K.}~\bibnamefont {Iida}}, \bibinfo {author}
  {\bibfnamefont {K.}~\bibnamefont {Kamazawa}}, \bibinfo {author}
  {\bibfnamefont {W.}~\bibnamefont {Yu}}, \bibinfo {author} {\bibfnamefont
  {X.}~\bibnamefont {Lin}},\ and\ \bibinfo {author} {\bibfnamefont
  {Y.}~\bibnamefont {Li}},\ }\bibfield  {title} {\bibinfo {title} {{Spin-orbit
  phase behavior of ${\mathrm{Na}}_{2}{\mathrm{Co}}_{2}{\mathrm{TeO}}_{6}$ at
  low temperatures}},\ }\href {https://doi.org/10.1103/PhysRevB.103.L180404}
  {\bibfield  {journal} {\bibinfo  {journal} {Phys. Rev. B}\ }\textbf {\bibinfo
  {volume} {103}},\ \bibinfo {pages} {L180404} (\bibinfo {year}
  {2021})}\BibitemShut {NoStop}%
\bibitem [{\citenamefont {Kim}\ \emph {et~al.}(2021{\natexlab{b}})\citenamefont
  {Kim}, \citenamefont {Jeong}, \citenamefont {Lin}, \citenamefont {Park},
  \citenamefont {Masuda}, \citenamefont {Asai}, \citenamefont {Itoh},
  \citenamefont {Kim}, \citenamefont {Zhou}, \citenamefont {Ma},\ and\
  \citenamefont {Park}}]{Kim2021b}%
  \BibitemOpen
  \bibfield  {author} {\bibinfo {author} {\bibfnamefont {C.}~\bibnamefont
  {Kim}}, \bibinfo {author} {\bibfnamefont {J.}~\bibnamefont {Jeong}}, \bibinfo
  {author} {\bibfnamefont {G.}~\bibnamefont {Lin}}, \bibinfo {author}
  {\bibfnamefont {P.}~\bibnamefont {Park}}, \bibinfo {author} {\bibfnamefont
  {T.}~\bibnamefont {Masuda}}, \bibinfo {author} {\bibfnamefont
  {S.}~\bibnamefont {Asai}}, \bibinfo {author} {\bibfnamefont {S.}~\bibnamefont
  {Itoh}}, \bibinfo {author} {\bibfnamefont {H.-S.}\ \bibnamefont {Kim}},
  \bibinfo {author} {\bibfnamefont {H.}~\bibnamefont {Zhou}}, \bibinfo {author}
  {\bibfnamefont {J.}~\bibnamefont {Ma}},\ and\ \bibinfo {author}
  {\bibfnamefont {J.-G.}\ \bibnamefont {Park}},\ }\bibfield  {title} {\bibinfo
  {title} {{Antiferromagnetic Kitaev interaction in $J_{\rm eff}=1/2$ cobalt
  honeycomb materials Na$_3$Co$_2$SbO$_6$ and Na$_2$Co$_2$TeO$_6$}},\ }\href
  {https://doi.org/10.1088/1361-648x/ac2644} {\bibfield  {journal} {\bibinfo
  {journal} {J. Condens. Matter Phys.}\ }\textbf {\bibinfo {volume} {34}},\
  \bibinfo {pages} {045802} (\bibinfo {year} {2021}{\natexlab{b}})}\BibitemShut
  {NoStop}%
\bibitem [{\citenamefont {Samarakoon}\ \emph {et~al.}(2021)\citenamefont
  {Samarakoon}, \citenamefont {Chen}, \citenamefont {Zhou},\ and\ \citenamefont
  {Garlea}}]{Samarakoon2021}%
  \BibitemOpen
  \bibfield  {author} {\bibinfo {author} {\bibfnamefont {A.~M.}\ \bibnamefont
  {Samarakoon}}, \bibinfo {author} {\bibfnamefont {Q.}~\bibnamefont {Chen}},
  \bibinfo {author} {\bibfnamefont {H.}~\bibnamefont {Zhou}},\ and\ \bibinfo
  {author} {\bibfnamefont {V.~O.}\ \bibnamefont {Garlea}},\ }\bibfield  {title}
  {\bibinfo {title} {{Static and dynamic magnetic properties of honeycomb
  lattice antiferromagnets ${\mathrm{Na}}_{2}{M}_{2}{\mathrm{TeO}}_{6}$,
  $M=\mathrm{Co}$ and Ni}},\ }\href
  {https://doi.org/10.1103/PhysRevB.104.184415} {\bibfield  {journal} {\bibinfo
   {journal} {Phys. Rev. B}\ }\textbf {\bibinfo {volume} {104}},\ \bibinfo
  {pages} {184415} (\bibinfo {year} {2021})}\BibitemShut {NoStop}%
\bibitem [{\citenamefont {Hong}\ \emph {et~al.}(2021)\citenamefont {Hong},
  \citenamefont {Gillig}, \citenamefont {Hentrich}, \citenamefont {Yao},
  \citenamefont {Kocsis}, \citenamefont {Witte}, \citenamefont {Schreiner},
  \citenamefont {Baumann}, \citenamefont {P\'erez}, \citenamefont {Wolter},
  \citenamefont {Li}, \citenamefont {B\"uchner},\ and\ \citenamefont
  {Hess}}]{Hong2021}%
  \BibitemOpen
  \bibfield  {author} {\bibinfo {author} {\bibfnamefont {X.}~\bibnamefont
  {Hong}}, \bibinfo {author} {\bibfnamefont {M.}~\bibnamefont {Gillig}},
  \bibinfo {author} {\bibfnamefont {R.}~\bibnamefont {Hentrich}}, \bibinfo
  {author} {\bibfnamefont {W.}~\bibnamefont {Yao}}, \bibinfo {author}
  {\bibfnamefont {V.}~\bibnamefont {Kocsis}}, \bibinfo {author} {\bibfnamefont
  {A.~R.}\ \bibnamefont {Witte}}, \bibinfo {author} {\bibfnamefont
  {T.}~\bibnamefont {Schreiner}}, \bibinfo {author} {\bibfnamefont
  {D.}~\bibnamefont {Baumann}}, \bibinfo {author} {\bibfnamefont
  {N.}~\bibnamefont {P\'erez}}, \bibinfo {author} {\bibfnamefont {A.~U.~B.}\
  \bibnamefont {Wolter}}, \bibinfo {author} {\bibfnamefont {Y.}~\bibnamefont
  {Li}}, \bibinfo {author} {\bibfnamefont {B.}~\bibnamefont {B\"uchner}},\ and\
  \bibinfo {author} {\bibfnamefont {C.}~\bibnamefont {Hess}},\ }\bibfield
  {title} {\bibinfo {title} {{Strongly scattered phonon heat transport of the
  candidate Kitaev material
  ${\mathrm{Na}}_{2}{\mathrm{Co}}_{2}{\mathrm{TeO}}_{6}$}},\ }\href
  {https://doi.org/10.1103/PhysRevB.104.144426} {\bibfield  {journal} {\bibinfo
   {journal} {Phys. Rev. B}\ }\textbf {\bibinfo {volume} {104}},\ \bibinfo
  {pages} {144426} (\bibinfo {year} {2021})}\BibitemShut {NoStop}%
\bibitem [{\citenamefont {Shirata}\ \emph {et~al.}(2012)\citenamefont
  {Shirata}, \citenamefont {Tanaka}, \citenamefont {Matsuo},\ and\
  \citenamefont {Kindo}}]{Shirata2012}%
  \BibitemOpen
  \bibfield  {author} {\bibinfo {author} {\bibfnamefont {Y.}~\bibnamefont
  {Shirata}}, \bibinfo {author} {\bibfnamefont {H.}~\bibnamefont {Tanaka}},
  \bibinfo {author} {\bibfnamefont {A.}~\bibnamefont {Matsuo}},\ and\ \bibinfo
  {author} {\bibfnamefont {K.}~\bibnamefont {Kindo}},\ }\bibfield  {title}
  {\bibinfo {title} {{Experimental Realization of a Spin-$1/2$
  Triangular-Lattice Heisenberg Antiferromagnet}},\ }\href
  {https://doi.org/10.1103/PhysRevLett.108.057205} {\bibfield  {journal}
  {\bibinfo  {journal} {Phys. Rev. Lett.}\ }\textbf {\bibinfo {volume} {108}},\
  \bibinfo {pages} {057205} (\bibinfo {year} {2012})}\BibitemShut {NoStop}%
\bibitem [{\citenamefont {Susuki}\ \emph {et~al.}(2013)\citenamefont {Susuki},
  \citenamefont {Kurita}, \citenamefont {Tanaka}, \citenamefont {Nojiri},
  \citenamefont {Matsuo}, \citenamefont {Kindo},\ and\ \citenamefont
  {Tanaka}}]{Susuki2013}%
  \BibitemOpen
  \bibfield  {author} {\bibinfo {author} {\bibfnamefont {T.}~\bibnamefont
  {Susuki}}, \bibinfo {author} {\bibfnamefont {N.}~\bibnamefont {Kurita}},
  \bibinfo {author} {\bibfnamefont {T.}~\bibnamefont {Tanaka}}, \bibinfo
  {author} {\bibfnamefont {H.}~\bibnamefont {Nojiri}}, \bibinfo {author}
  {\bibfnamefont {A.}~\bibnamefont {Matsuo}}, \bibinfo {author} {\bibfnamefont
  {K.}~\bibnamefont {Kindo}},\ and\ \bibinfo {author} {\bibfnamefont
  {H.}~\bibnamefont {Tanaka}},\ }\bibfield  {title} {\bibinfo {title}
  {{Magnetization Process and Collective Excitations in the $S\mathbf{=}1/2$
  Triangular-Lattice Heisenberg Antiferromagnet
  ${\mathrm{Ba}}_{3}{\mathrm{CoSb}}_{2}{\mathrm{O}}_{9}$}},\ }\href
  {https://doi.org/10.1103/PhysRevLett.110.267201} {\bibfield  {journal}
  {\bibinfo  {journal} {Phys. Rev. Lett.}\ }\textbf {\bibinfo {volume} {110}},\
  \bibinfo {pages} {267201} (\bibinfo {year} {2013})}\BibitemShut {NoStop}%
\bibitem [{\citenamefont {Koutroulakis}\ \emph {et~al.}(2015)\citenamefont
  {Koutroulakis}, \citenamefont {Zhou}, \citenamefont {Kamiya}, \citenamefont
  {Thompson}, \citenamefont {Zhou}, \citenamefont {Batista},\ and\
  \citenamefont {Brown}}]{Koutroulakis2015}%
  \BibitemOpen
  \bibfield  {author} {\bibinfo {author} {\bibfnamefont {G.}~\bibnamefont
  {Koutroulakis}}, \bibinfo {author} {\bibfnamefont {T.}~\bibnamefont {Zhou}},
  \bibinfo {author} {\bibfnamefont {Y.}~\bibnamefont {Kamiya}}, \bibinfo
  {author} {\bibfnamefont {J.~D.}\ \bibnamefont {Thompson}}, \bibinfo {author}
  {\bibfnamefont {H.~D.}\ \bibnamefont {Zhou}}, \bibinfo {author}
  {\bibfnamefont {C.~D.}\ \bibnamefont {Batista}},\ and\ \bibinfo {author}
  {\bibfnamefont {S.~E.}\ \bibnamefont {Brown}},\ }\bibfield  {title} {\bibinfo
  {title} {{Quantum phase diagram of the $S=\frac{1}{2}$ triangular-lattice
  antiferromagnet ${\mathrm{Ba}}_{3}{\mathrm{CoSb}}_{2}{\mathrm{O}}_{9}$}},\
  }\href {https://doi.org/10.1103/PhysRevB.91.024410} {\bibfield  {journal}
  {\bibinfo  {journal} {Phys. Rev. B}\ }\textbf {\bibinfo {volume} {91}},\
  \bibinfo {pages} {024410} (\bibinfo {year} {2015})}\BibitemShut {NoStop}%
\bibitem [{\citenamefont {Zhong}\ \emph {et~al.}(2019)\citenamefont {Zhong},
  \citenamefont {Guo}, \citenamefont {Xu}, \citenamefont {Xu},\ and\
  \citenamefont {Cava}}]{Zhong19}%
  \BibitemOpen
  \bibfield  {author} {\bibinfo {author} {\bibfnamefont {R.}~\bibnamefont
  {Zhong}}, \bibinfo {author} {\bibfnamefont {S.}~\bibnamefont {Guo}}, \bibinfo
  {author} {\bibfnamefont {G.}~\bibnamefont {Xu}}, \bibinfo {author}
  {\bibfnamefont {Z.}~\bibnamefont {Xu}},\ and\ \bibinfo {author}
  {\bibfnamefont {R.~J.}\ \bibnamefont {Cava}},\ }\bibfield  {title} {\bibinfo
  {title} {{Strong quantum fluctuations in a quantum spin liquid candidate with
  a {Co}-based triangular lattice}},\ }\href
  {https://doi.org/10.1073/pnas.1906483116} {\bibfield  {journal} {\bibinfo
  {journal} {Proc. Nat. Acad. Sci.}\ }\textbf {\bibinfo {volume} {116}},\
  \bibinfo {pages} {14505} (\bibinfo {year} {2019})}\BibitemShut {NoStop}%
\bibitem [{\citenamefont {Li}\ \emph {et~al.}(2020)\citenamefont {Li},
  \citenamefont {Huang}, \citenamefont {Yue}, \citenamefont {Chu},
  \citenamefont {Chen}, \citenamefont {Choi}, \citenamefont {Zhao},
  \citenamefont {Zhou},\ and\ \citenamefont {Sun}}]{Li2020}%
  \BibitemOpen
  \bibfield  {author} {\bibinfo {author} {\bibfnamefont {N.}~\bibnamefont
  {Li}}, \bibinfo {author} {\bibfnamefont {Q.}~\bibnamefont {Huang}}, \bibinfo
  {author} {\bibfnamefont {X.~Y.}\ \bibnamefont {Yue}}, \bibinfo {author}
  {\bibfnamefont {W.~J.}\ \bibnamefont {Chu}}, \bibinfo {author} {\bibfnamefont
  {Q.}~\bibnamefont {Chen}}, \bibinfo {author} {\bibfnamefont {E.~S.}\
  \bibnamefont {Choi}}, \bibinfo {author} {\bibfnamefont {X.}~\bibnamefont
  {Zhao}}, \bibinfo {author} {\bibfnamefont {H.~D.}\ \bibnamefont {Zhou}},\
  and\ \bibinfo {author} {\bibfnamefont {X.~F.}\ \bibnamefont {Sun}},\
  }\bibfield  {title} {\bibinfo {title} {Possible itinerant excitations and
  quantum spinstate transitions in the effective spin-1/2 triangular-lattice
  antiferromagnet {Na$_2$BaCo(PO$_4$)$_2$}},\ }\href
  {https://doi.org/10.1038/s41467-020-18041-3} {\bibfield  {journal} {\bibinfo
  {journal} {Nat. Commun.}\ }\textbf {\bibinfo {volume} {11}},\ \bibinfo
  {pages} {4216} (\bibinfo {year} {2020})}\BibitemShut {NoStop}%
\bibitem [{\citenamefont {{Lee, S. and Lee, C. H. and Berlie, A. and Hillier,
  A. D. and Adroja, Devashibhai T. and Zhong, Ruidan and Cava, R. J. and Jang,
  Z. H. and Choi, K.-Y.}}(2021)}]{Lee2021}%
  \BibitemOpen
  \bibfield  {author} {\bibinfo {author} {\bibnamefont {{Lee, S. and Lee, C. H.
  and Berlie, A. and Hillier, A. D. and Adroja, Devashibhai T. and Zhong,
  Ruidan and Cava, R. J. and Jang, Z. H. and Choi, K.-Y.}}},\ }\bibfield
  {title} {\bibinfo {title} {{Temporal and field evolution of spin excitations
  in the disorder-free triangular antiferromagnet
  ${\mathrm{Na}}_{2}\mathrm{BaCo}{({\mathrm{PO}}_{4})}_{2}$}},\ }\href
  {https://doi.org/10.1103/PhysRevB.103.024413} {\bibfield  {journal} {\bibinfo
   {journal} {{Phys. Rev. B}}\ }\textbf {\bibinfo {volume} {103}},\ \bibinfo
  {pages} {024413} (\bibinfo {year} {2021})}\BibitemShut {NoStop}%
\bibitem [{\citenamefont {Wellm}\ \emph {et~al.}(2021)\citenamefont {Wellm},
  \citenamefont {Roscher}, \citenamefont {Zeisner}, \citenamefont {Alfonsov},
  \citenamefont {Zhong}, \citenamefont {Cava}, \citenamefont {Savoyant},
  \citenamefont {Hayn}, \citenamefont {van~den Brink}, \citenamefont
  {B\"uchner}, \citenamefont {Janson},\ and\ \citenamefont
  {Kataev}}]{Wellm2021}%
  \BibitemOpen
  \bibfield  {author} {\bibinfo {author} {\bibfnamefont {C.}~\bibnamefont
  {Wellm}}, \bibinfo {author} {\bibfnamefont {W.}~\bibnamefont {Roscher}},
  \bibinfo {author} {\bibfnamefont {J.}~\bibnamefont {Zeisner}}, \bibinfo
  {author} {\bibfnamefont {A.}~\bibnamefont {Alfonsov}}, \bibinfo {author}
  {\bibfnamefont {R.}~\bibnamefont {Zhong}}, \bibinfo {author} {\bibfnamefont
  {R.~J.}\ \bibnamefont {Cava}}, \bibinfo {author} {\bibfnamefont
  {A.}~\bibnamefont {Savoyant}}, \bibinfo {author} {\bibfnamefont
  {R.}~\bibnamefont {Hayn}}, \bibinfo {author} {\bibfnamefont {J.}~\bibnamefont
  {van~den Brink}}, \bibinfo {author} {\bibfnamefont {B.}~\bibnamefont
  {B\"uchner}}, \bibinfo {author} {\bibfnamefont {O.}~\bibnamefont {Janson}},\
  and\ \bibinfo {author} {\bibfnamefont {V.}~\bibnamefont {Kataev}},\
  }\bibfield  {title} {\bibinfo {title} {{Frustration enhanced by Kitaev
  exchange in a ${\stackrel{\ifmmode \tilde{}\else
  \~{}\fi{}}{j}}_{\text{eff}}=\frac{1}{2}$ triangular antiferromagnet}},\
  }\href {https://doi.org/10.1103/PhysRevB.104.L100420} {\bibfield  {journal}
  {\bibinfo  {journal} {Phys. Rev. B}\ }\textbf {\bibinfo {volume} {104}},\
  \bibinfo {pages} {L100420} (\bibinfo {year} {2021})}\BibitemShut {NoStop}%
\bibitem [{\citenamefont {Tanaka}\ \emph {et~al.}(2001)\citenamefont {Tanaka},
  \citenamefont {Oosawa}, \citenamefont {Kato}, \citenamefont {Uekusa},
  \citenamefont {Ohashi}, \citenamefont {Kakurai},\ and\ \citenamefont
  {Hoser}}]{Tanaka2001}%
  \BibitemOpen
  \bibfield  {author} {\bibinfo {author} {\bibfnamefont {H.}~\bibnamefont
  {Tanaka}}, \bibinfo {author} {\bibfnamefont {A.}~\bibnamefont {Oosawa}},
  \bibinfo {author} {\bibfnamefont {T.}~\bibnamefont {Kato}}, \bibinfo {author}
  {\bibfnamefont {H.}~\bibnamefont {Uekusa}}, \bibinfo {author} {\bibfnamefont
  {Y.}~\bibnamefont {Ohashi}}, \bibinfo {author} {\bibfnamefont
  {K.}~\bibnamefont {Kakurai}},\ and\ \bibinfo {author} {\bibfnamefont
  {A.}~\bibnamefont {Hoser}},\ }\bibfield  {title} {\bibinfo {title}
  {{Observation of field-induced transverse N{\'e}el ordering in the spin gap
  system TlCuCl$_3$}},\ }\href {https://doi.org/10.1143/JPSJ.70.939} {\bibfield
   {journal} {\bibinfo  {journal} {J. Phys. Soc. Japan}\ }\textbf {\bibinfo
  {volume} {70}},\ \bibinfo {pages} {939} (\bibinfo {year} {2001})}\BibitemShut
  {NoStop}%
\bibitem [{\citenamefont {Masuda}\ \emph {et~al.}(2006)\citenamefont {Masuda},
  \citenamefont {Zheludev}, \citenamefont {Manaka}, \citenamefont {Regnault},
  \citenamefont {Chung},\ and\ \citenamefont {Qiu}}]{Masuda2006}%
  \BibitemOpen
  \bibfield  {author} {\bibinfo {author} {\bibfnamefont {T.}~\bibnamefont
  {Masuda}}, \bibinfo {author} {\bibfnamefont {A.}~\bibnamefont {Zheludev}},
  \bibinfo {author} {\bibfnamefont {H.}~\bibnamefont {Manaka}}, \bibinfo
  {author} {\bibfnamefont {L.-P.}\ \bibnamefont {Regnault}}, \bibinfo {author}
  {\bibfnamefont {J.-H.}\ \bibnamefont {Chung}},\ and\ \bibinfo {author}
  {\bibfnamefont {Y.}~\bibnamefont {Qiu}},\ }\bibfield  {title} {\bibinfo
  {title} {{Dynamics of Composite Haldane Spin Chains in IPA--CuCl$_4$}},\
  }\href {https://doi.org/10.1103/PhysRevLett.96.047210} {\bibfield  {journal}
  {\bibinfo  {journal} {Phys. Rev. Lett.}\ }\textbf {\bibinfo {volume} {96}},\
  \bibinfo {pages} {047210} (\bibinfo {year} {2006})}\BibitemShut {NoStop}%
\bibitem [{\citenamefont {Garlea}\ \emph {et~al.}(2007)\citenamefont {Garlea},
  \citenamefont {Zheludev}, \citenamefont {Masuda}, \citenamefont {Manaka},
  \citenamefont {Regnault}, \citenamefont {Ressouche}, \citenamefont {Grenier},
  \citenamefont {Chung}, \citenamefont {Qiu}, \citenamefont {Habicht},
  \citenamefont {Kiefer},\ and\ \citenamefont {Boehm}}]{Garlea2007}%
  \BibitemOpen
  \bibfield  {author} {\bibinfo {author} {\bibfnamefont {V.~O.}\ \bibnamefont
  {Garlea}}, \bibinfo {author} {\bibfnamefont {A.}~\bibnamefont {Zheludev}},
  \bibinfo {author} {\bibfnamefont {T.}~\bibnamefont {Masuda}}, \bibinfo
  {author} {\bibfnamefont {H.}~\bibnamefont {Manaka}}, \bibinfo {author}
  {\bibfnamefont {L.-P.}\ \bibnamefont {Regnault}}, \bibinfo {author}
  {\bibfnamefont {E.}~\bibnamefont {Ressouche}}, \bibinfo {author}
  {\bibfnamefont {B.}~\bibnamefont {Grenier}}, \bibinfo {author} {\bibfnamefont
  {J.-H.}\ \bibnamefont {Chung}}, \bibinfo {author} {\bibfnamefont
  {Y.}~\bibnamefont {Qiu}}, \bibinfo {author} {\bibfnamefont {K.}~\bibnamefont
  {Habicht}}, \bibinfo {author} {\bibfnamefont {K.}~\bibnamefont {Kiefer}},\
  and\ \bibinfo {author} {\bibfnamefont {M.}~\bibnamefont {Boehm}},\ }\bibfield
   {title} {\bibinfo {title} {{Excitations from a Bose-Einstein Condensate of
  Magnons in Coupled Spin Ladders}},\ }\href
  {https://doi.org/10.1103/PhysRevLett.98.167202} {\bibfield  {journal}
  {\bibinfo  {journal} {Phys. Rev. Lett.}\ }\textbf {\bibinfo {volume} {98}},\
  \bibinfo {pages} {167202} (\bibinfo {year} {2007})}\BibitemShut {NoStop}%
\bibitem [{\citenamefont {Klanj\ifmmode~\check{s}\else \v{s}\fi{}ek}\ \emph
  {et~al.}(2008)\citenamefont {Klanj\ifmmode~\check{s}\else \v{s}\fi{}ek},
  \citenamefont {Mayaffre}, \citenamefont {Berthier}, \citenamefont
  {Horvati\ifmmode~\acute{c}\else \'{c}\fi{}}, \citenamefont {Chiari},
  \citenamefont {Piovesana}, \citenamefont {Bouillot}, \citenamefont {Kollath},
  \citenamefont {Orignac}, \citenamefont {Citro},\ and\ \citenamefont
  {Giamarchi}}]{Klanjsek2008}%
  \BibitemOpen
  \bibfield  {author} {\bibinfo {author} {\bibfnamefont {M.}~\bibnamefont
  {Klanj\ifmmode~\check{s}\else \v{s}\fi{}ek}}, \bibinfo {author}
  {\bibfnamefont {H.}~\bibnamefont {Mayaffre}}, \bibinfo {author}
  {\bibfnamefont {C.}~\bibnamefont {Berthier}}, \bibinfo {author}
  {\bibfnamefont {M.}~\bibnamefont {Horvati\ifmmode~\acute{c}\else
  \'{c}\fi{}}}, \bibinfo {author} {\bibfnamefont {B.}~\bibnamefont {Chiari}},
  \bibinfo {author} {\bibfnamefont {O.}~\bibnamefont {Piovesana}}, \bibinfo
  {author} {\bibfnamefont {P.}~\bibnamefont {Bouillot}}, \bibinfo {author}
  {\bibfnamefont {C.}~\bibnamefont {Kollath}}, \bibinfo {author} {\bibfnamefont
  {E.}~\bibnamefont {Orignac}}, \bibinfo {author} {\bibfnamefont
  {R.}~\bibnamefont {Citro}},\ and\ \bibinfo {author} {\bibfnamefont
  {T.}~\bibnamefont {Giamarchi}},\ }\bibfield  {title} {\bibinfo {title}
  {{Controlling Luttinger Liquid Physics in Spin Ladders under a Magnetic
  Field}},\ }\href {https://doi.org/10.1103/PhysRevLett.101.137207} {\bibfield
  {journal} {\bibinfo  {journal} {Phys. Rev. Lett.}\ }\textbf {\bibinfo
  {volume} {101}},\ \bibinfo {pages} {137207} (\bibinfo {year}
  {2008})}\BibitemShut {NoStop}%
\bibitem [{\citenamefont {Thielemann}\ \emph {et~al.}(2009)\citenamefont
  {Thielemann}, \citenamefont {R\"uegg}, \citenamefont {R\o{}nnow},
  \citenamefont {L\"auchli}, \citenamefont {Caux}, \citenamefont {Normand},
  \citenamefont {Biner}, \citenamefont {Kr\"amer}, \citenamefont {G\"udel},
  \citenamefont {Stahn}, \citenamefont {Habicht}, \citenamefont {Kiefer},
  \citenamefont {Boehm}, \citenamefont {McMorrow},\ and\ \citenamefont
  {Mesot}}]{Thielemann2009}%
  \BibitemOpen
  \bibfield  {author} {\bibinfo {author} {\bibfnamefont {B.}~\bibnamefont
  {Thielemann}}, \bibinfo {author} {\bibfnamefont {C.}~\bibnamefont {R\"uegg}},
  \bibinfo {author} {\bibfnamefont {H.~M.}\ \bibnamefont {R\o{}nnow}}, \bibinfo
  {author} {\bibfnamefont {A.~M.}\ \bibnamefont {L\"auchli}}, \bibinfo {author}
  {\bibfnamefont {J.-S.}\ \bibnamefont {Caux}}, \bibinfo {author}
  {\bibfnamefont {B.}~\bibnamefont {Normand}}, \bibinfo {author} {\bibfnamefont
  {D.}~\bibnamefont {Biner}}, \bibinfo {author} {\bibfnamefont {K.~W.}\
  \bibnamefont {Kr\"amer}}, \bibinfo {author} {\bibfnamefont {H.-U.}\
  \bibnamefont {G\"udel}}, \bibinfo {author} {\bibfnamefont {J.}~\bibnamefont
  {Stahn}}, \bibinfo {author} {\bibfnamefont {K.}~\bibnamefont {Habicht}},
  \bibinfo {author} {\bibfnamefont {K.}~\bibnamefont {Kiefer}}, \bibinfo
  {author} {\bibfnamefont {M.}~\bibnamefont {Boehm}}, \bibinfo {author}
  {\bibfnamefont {D.~F.}\ \bibnamefont {McMorrow}},\ and\ \bibinfo {author}
  {\bibfnamefont {J.}~\bibnamefont {Mesot}},\ }\bibfield  {title} {\bibinfo
  {title} {{Direct Observation of Magnon Fractionalization in the Quantum Spin
  Ladder}},\ }\href {https://doi.org/10.1103/PhysRevLett.102.107204} {\bibfield
   {journal} {\bibinfo  {journal} {Phys. Rev. Lett.}\ }\textbf {\bibinfo
  {volume} {102}},\ \bibinfo {pages} {107204} (\bibinfo {year}
  {2009})}\BibitemShut {NoStop}%
\bibitem [{\citenamefont {Zapf}\ \emph {et~al.}(2014)\citenamefont {Zapf},
  \citenamefont {Jaime},\ and\ \citenamefont {Batista}}]{Zapf2014}%
  \BibitemOpen
  \bibfield  {author} {\bibinfo {author} {\bibfnamefont {V.}~\bibnamefont
  {Zapf}}, \bibinfo {author} {\bibfnamefont {M.}~\bibnamefont {Jaime}},\ and\
  \bibinfo {author} {\bibfnamefont {C.~D.}\ \bibnamefont {Batista}},\
  }\bibfield  {title} {\bibinfo {title} {{Bose-Einstein condensation in quantum
  magnets}},\ }\href {https://doi.org/10.1103/RevModPhys.86.563} {\bibfield
  {journal} {\bibinfo  {journal} {Rev. Mod. Phys.}\ }\textbf {\bibinfo {volume}
  {86}},\ \bibinfo {pages} {563} (\bibinfo {year} {2014})}\BibitemShut
  {NoStop}%
\bibitem [{\citenamefont {Nadir}\ \emph {et~al.}(1999)\citenamefont {Nadir},
  \citenamefont {Swinnea},\ and\ \citenamefont {Steinfink}}]{Nadir1999}%
  \BibitemOpen
  \bibfield  {author} {\bibinfo {author} {\bibfnamefont {S.}~\bibnamefont
  {Nadir}}, \bibinfo {author} {\bibfnamefont {J.}~\bibnamefont {Swinnea}},\
  and\ \bibinfo {author} {\bibfnamefont {H.}~\bibnamefont {Steinfink}},\
  }\bibfield  {title} {\bibinfo {title} {Crystal structure and physical
  properties of {BiCoPO$_5$}},\ }\href
  {https://doi.org/https://doi.org/10.1006/jssc.1999.8450} {\bibfield
  {journal} {\bibinfo  {journal} {J. Solid State Chem.}\ }\textbf {\bibinfo
  {volume} {148}},\ \bibinfo {pages} {295} (\bibinfo {year}
  {1999})}\BibitemShut {NoStop}%
\bibitem [{\citenamefont {Ketatni}\ \emph {et~al.}(1999)\citenamefont
  {Ketatni}, \citenamefont {Abraham},\ and\ \citenamefont
  {Mentre}}]{Ketatni1999}%
  \BibitemOpen
  \bibfield  {author} {\bibinfo {author} {\bibfnamefont {M.}~\bibnamefont
  {Ketatni}}, \bibinfo {author} {\bibfnamefont {F.}~\bibnamefont {Abraham}},\
  and\ \bibinfo {author} {\bibfnamefont {O.}~\bibnamefont {Mentre}},\
  }\bibfield  {title} {\bibinfo {title} {{Channel structure in the new
  BiCoPO$_5$. Comparison with BiNiPO$_5$. Crystal structure, lone pair
  localisation and infrared characterisation}},\ }\href
  {https://doi.org/https://doi.org/10.1016/S1293-2558(00)80097-7} {\bibfield
  {journal} {\bibinfo  {journal} {Solid State Sc.}\ }\textbf {\bibinfo {volume}
  {1}},\ \bibinfo {pages} {449} (\bibinfo {year} {1999})}\BibitemShut {NoStop}%
\bibitem [{\citenamefont {Mentre}\ \emph {et~al.}(2008)\citenamefont {Mentre},
  \citenamefont {Bouree}, \citenamefont {Rodriguez-Carvajal}, \citenamefont
  {Jazouli}, \citenamefont {Khayati},\ and\ \citenamefont
  {Ketatni}}]{Mentre2008}%
  \BibitemOpen
  \bibfield  {author} {\bibinfo {author} {\bibfnamefont {O.}~\bibnamefont
  {Mentre}}, \bibinfo {author} {\bibfnamefont {F.}~\bibnamefont {Bouree}},
  \bibinfo {author} {\bibfnamefont {J.}~\bibnamefont {Rodriguez-Carvajal}},
  \bibinfo {author} {\bibfnamefont {A.~E.}\ \bibnamefont {Jazouli}}, \bibinfo
  {author} {\bibfnamefont {N.~E.}\ \bibnamefont {Khayati}},\ and\ \bibinfo
  {author} {\bibfnamefont {E.~M.}\ \bibnamefont {Ketatni}},\ }\bibfield
  {title} {\bibinfo {title} {Magnetic structure and analysis of the exchange
  interactions in {BiMO}({PO}$_4$) {(M = Co, Ni)}},\ }\href
  {https://doi.org/10.1088/0953-8984/20/41/415211} {\bibfield  {journal}
  {\bibinfo  {journal} {J. Phys.: Condens. Matter}\ }\textbf {\bibinfo {volume}
  {20}},\ \bibinfo {pages} {415211} (\bibinfo {year} {2008})}\BibitemShut
  {NoStop}%
\bibitem [{\citenamefont {Mathews}\ \emph {et~al.}(2013)\citenamefont
  {Mathews}, \citenamefont {Ranjith}, \citenamefont {Baenitz},\ and\
  \citenamefont {Nath}}]{Mathews2013}%
  \BibitemOpen
  \bibfield  {author} {\bibinfo {author} {\bibfnamefont {E.}~\bibnamefont
  {Mathews}}, \bibinfo {author} {\bibfnamefont {K.}~\bibnamefont {Ranjith}},
  \bibinfo {author} {\bibfnamefont {M.}~\bibnamefont {Baenitz}},\ and\ \bibinfo
  {author} {\bibfnamefont {R.}~\bibnamefont {Nath}},\ }\bibfield  {title}
  {\bibinfo {title} {{Field induced magnetic transition in low-dimensional
  magnets Bi(Ni,Co)PO$_5$}},\ }\href
  {https://doi.org/https://doi.org/10.1016/j.ssc.2012.10.033} {\bibfield
  {journal} {\bibinfo  {journal} {Solid State Commun.}\ }\textbf {\bibinfo
  {volume} {154}},\ \bibinfo {pages} {56} (\bibinfo {year} {2013})}\BibitemShut
  {NoStop}%
\bibitem [{\citenamefont {Cai}\ \emph {et~al.}(2015)\citenamefont {Cai},
  \citenamefont {Yang},\ and\ \citenamefont {Kusmartsev}}]{Cai2015}%
  \BibitemOpen
  \bibfield  {author} {\bibinfo {author} {\bibfnamefont {R.-G.}\ \bibnamefont
  {Cai}}, \bibinfo {author} {\bibfnamefont {R.-Q.}\ \bibnamefont {Yang}},\ and\
  \bibinfo {author} {\bibfnamefont {F.~V.}\ \bibnamefont {Kusmartsev}},\
  }\bibfield  {title} {\bibinfo {title} {{Holographic antiferromagnetic quantum
  criticality and ${\mathrm{AdS}}_{2}$ scaling limit}},\ }\href
  {https://doi.org/10.1103/PhysRevD.92.046005} {\bibfield  {journal} {\bibinfo
  {journal} {Phys. Rev. D}\ }\textbf {\bibinfo {volume} {92}},\ \bibinfo
  {pages} {046005} (\bibinfo {year} {2015})}\BibitemShut {NoStop}%
\bibitem [{\citenamefont {Skourski}\ \emph {et~al.}(2011)\citenamefont
  {Skourski}, \citenamefont {Kuz'min}, \citenamefont {Skokov}, \citenamefont
  {Andreev},\ and\ \citenamefont {Wosnitza}}]{Skourski2011}%
  \BibitemOpen
  \bibfield  {author} {\bibinfo {author} {\bibfnamefont {Y.}~\bibnamefont
  {Skourski}}, \bibinfo {author} {\bibfnamefont {M.~D.}\ \bibnamefont
  {Kuz'min}}, \bibinfo {author} {\bibfnamefont {K.~P.}\ \bibnamefont {Skokov}},
  \bibinfo {author} {\bibfnamefont {A.~V.}\ \bibnamefont {Andreev}},\ and\
  \bibinfo {author} {\bibfnamefont {J.}~\bibnamefont {Wosnitza}},\ }\bibfield
  {title} {\bibinfo {title} {{High-field magnetization of
  Ho${}_{2}$Fe${}_{17}$}},\ }\href {https://doi.org/10.1103/PhysRevB.83.214420}
  {\bibfield  {journal} {\bibinfo  {journal} {Phys. Rev. B}\ }\textbf {\bibinfo
  {volume} {83}},\ \bibinfo {pages} {214420} (\bibinfo {year}
  {2011})}\BibitemShut {NoStop}%
\bibitem [{\citenamefont {Alfonsov}\ \emph {et~al.}(2021)\citenamefont
  {Alfonsov}, \citenamefont {Mehlawat}, \citenamefont {Zeugner}, \citenamefont
  {Isaeva}, \citenamefont {B\"uchner},\ and\ \citenamefont
  {Kataev}}]{Alfonsov2021}%
  \BibitemOpen
  \bibfield  {author} {\bibinfo {author} {\bibfnamefont {A.}~\bibnamefont
  {Alfonsov}}, \bibinfo {author} {\bibfnamefont {K.}~\bibnamefont {Mehlawat}},
  \bibinfo {author} {\bibfnamefont {A.}~\bibnamefont {Zeugner}}, \bibinfo
  {author} {\bibfnamefont {A.}~\bibnamefont {Isaeva}}, \bibinfo {author}
  {\bibfnamefont {B.}~\bibnamefont {B\"uchner}},\ and\ \bibinfo {author}
  {\bibfnamefont {V.}~\bibnamefont {Kataev}},\ }\bibfield  {title} {\bibinfo
  {title} {{Magnetic-field tuning of the spin dynamics in the magnetic
  topological insulators
  $({\mathrm{MnBi}}_{2}{\mathrm{Te}}_{4}){({\mathrm{Bi}}_{2}{\mathrm{Te}}_{3})}_{n}$}},\
  }\href {https://doi.org/10.1103/PhysRevB.104.195139} {\bibfield  {journal}
  {\bibinfo  {journal} {Phys. Rev. B}\ }\textbf {\bibinfo {volume} {104}},\
  \bibinfo {pages} {195139} (\bibinfo {year} {2021})}\BibitemShut {NoStop}%
\bibitem [{\citenamefont {Derenzo}\ \emph {et~al.}(2000)\citenamefont
  {Derenzo}, \citenamefont {Klintenberg},\ and\ \citenamefont
  {Weber}}]{Derenzo2000}%
  \BibitemOpen
  \bibfield  {author} {\bibinfo {author} {\bibfnamefont {S.~E.}\ \bibnamefont
  {Derenzo}}, \bibinfo {author} {\bibfnamefont {M.~K.}\ \bibnamefont
  {Klintenberg}},\ and\ \bibinfo {author} {\bibfnamefont {M.~J.}\ \bibnamefont
  {Weber}},\ }\bibfield  {title} {\bibinfo {title} {Determining point charge
  arrays that produce accurate ionic crystal fields for atomic cluster
  calculations},\ }\href {https://doi.org/10.1063/1.480776} {\bibfield
  {journal} {\bibinfo  {journal} {J. Chem. Phys.}\ }\textbf {\bibinfo {volume}
  {112}},\ \bibinfo {pages} {2074} (\bibinfo {year} {2000})}\BibitemShut
  {NoStop}%
\bibitem [{\citenamefont {Klintenberg}\ \emph {et~al.}(2000)\citenamefont
  {Klintenberg}, \citenamefont {Derenzo},\ and\ \citenamefont
  {Weber}}]{Klintenberg2000}%
  \BibitemOpen
  \bibfield  {author} {\bibinfo {author} {\bibfnamefont {M.}~\bibnamefont
  {Klintenberg}}, \bibinfo {author} {\bibfnamefont {S.}~\bibnamefont
  {Derenzo}},\ and\ \bibinfo {author} {\bibfnamefont {M.}~\bibnamefont
  {Weber}},\ }\bibfield  {title} {\bibinfo {title} {Accurate crystal fields for
  embedded cluster calculations},\ }\href
  {https://doi.org/10.1016/S0010-4655(00)00071-0} {\bibfield  {journal}
  {\bibinfo  {journal} {Comp. Phys. Commun.}\ }\textbf {\bibinfo {volume}
  {131}},\ \bibinfo {pages} {120} (\bibinfo {year} {2000})}\BibitemShut
  {NoStop}%
\bibitem [{\citenamefont {Küchle}\ \emph {et~al.}(1991)\citenamefont
  {Küchle}, \citenamefont {Dolg}, \citenamefont {Stoll},\ and\ \citenamefont
  {Preuss}}]{Kuechle1991}%
  \BibitemOpen
  \bibfield  {author} {\bibinfo {author} {\bibfnamefont {W.}~\bibnamefont
  {Küchle}}, \bibinfo {author} {\bibfnamefont {M.}~\bibnamefont {Dolg}},
  \bibinfo {author} {\bibfnamefont {H.}~\bibnamefont {Stoll}},\ and\ \bibinfo
  {author} {\bibfnamefont {H.}~\bibnamefont {Preuss}},\ }\bibfield  {title}
  {\bibinfo {title} {Ab initio pseudopotentials for {Hg} through {Rn}},\ }\href
  {https://doi.org/10.1080/00268979100102941} {\bibfield  {journal} {\bibinfo
  {journal} {Mol. Phys.}\ }\textbf {\bibinfo {volume} {74}},\ \bibinfo {pages}
  {1245} (\bibinfo {year} {1991})}\BibitemShut {NoStop}%
\bibitem [{\citenamefont {Dolg}\ \emph {et~al.}(1987)\citenamefont {Dolg},
  \citenamefont {Wedig}, \citenamefont {Stoll},\ and\ \citenamefont
  {Preuss}}]{Dolg1987}%
  \BibitemOpen
  \bibfield  {author} {\bibinfo {author} {\bibfnamefont {M.}~\bibnamefont
  {Dolg}}, \bibinfo {author} {\bibfnamefont {U.}~\bibnamefont {Wedig}},
  \bibinfo {author} {\bibfnamefont {H.}~\bibnamefont {Stoll}},\ and\ \bibinfo
  {author} {\bibfnamefont {H.}~\bibnamefont {Preuss}},\ }\bibfield  {title}
  {\bibinfo {title} {Energy‐adjusted ab initio pseudopotentials for the first
  row transition elements},\ }\href {https://doi.org/10.1063/1.452288}
  {\bibfield  {journal} {\bibinfo  {journal} {J. Chem. Phys.}\ }\textbf
  {\bibinfo {volume} {86}},\ \bibinfo {pages} {866} (\bibinfo {year}
  {1987})}\BibitemShut {NoStop}%
\bibitem [{\citenamefont {Weigend}\ and\ \citenamefont
  {Ahlrichs}(2005)}]{Weigend2005}%
  \BibitemOpen
  \bibfield  {author} {\bibinfo {author} {\bibfnamefont {F.}~\bibnamefont
  {Weigend}}\ and\ \bibinfo {author} {\bibfnamefont {R.}~\bibnamefont
  {Ahlrichs}},\ }\bibfield  {title} {\bibinfo {title} {Balanced basis sets of
  split valence{,} triple zeta valence and quadruple zeta valence quality for
  {H} to {Rn}: Design and assessment of accuracy},\ }\href
  {https://doi.org/10.1039/B508541A} {\bibfield  {journal} {\bibinfo  {journal}
  {Phys. Chem. Chem. Phys.}\ }\textbf {\bibinfo {volume} {7}},\ \bibinfo
  {pages} {3297} (\bibinfo {year} {2005})}\BibitemShut {NoStop}%
\bibitem [{\citenamefont {Roos}(1987)}]{Roos1987}%
  \BibitemOpen
  \bibfield  {author} {\bibinfo {author} {\bibfnamefont {B.~O.}\ \bibnamefont
  {Roos}},\ }\bibinfo {title} {The complete active space self-consistent field
  method and its applications in electronic structure calculations},\ in\ \href
  {https://doi.org/10.1002/9780470142943.ch7} {\emph {\bibinfo {booktitle}
  {Adv. Chem. Phys.}}}\ (\bibinfo  {publisher} {John Wiley \& Sons, Ltd},\
  \bibinfo {year} {1987})\ pp.\ \bibinfo {pages} {399--445}\BibitemShut
  {NoStop}%
\bibitem [{\citenamefont {Douglas}\ and\ \citenamefont
  {Kroll}(1974)}]{Douglas1974}%
  \BibitemOpen
  \bibfield  {author} {\bibinfo {author} {\bibfnamefont {M.}~\bibnamefont
  {Douglas}}\ and\ \bibinfo {author} {\bibfnamefont {N.~M.}\ \bibnamefont
  {Kroll}},\ }\bibfield  {title} {\bibinfo {title} {Quantum electrodynamical
  corrections to the fine structure of helium},\ }\href
  {https://doi.org/10.1016/0003-4916(74)90333-9} {\bibfield  {journal}
  {\bibinfo  {journal} {Ann. Phys.}\ }\textbf {\bibinfo {volume} {82}},\
  \bibinfo {pages} {89} (\bibinfo {year} {1974})}\BibitemShut {NoStop}%
\bibitem [{\citenamefont {Hess}(1986)}]{Hess1986}%
  \BibitemOpen
  \bibfield  {author} {\bibinfo {author} {\bibfnamefont {B.~A.}\ \bibnamefont
  {Hess}},\ }\bibfield  {title} {\bibinfo {title} {Relativistic
  electronic-structure calculations employing a two-component no-pair formalism
  with external-field projection operators},\ }\href
  {https://doi.org/10.1103/PhysRevA.33.3742} {\bibfield  {journal} {\bibinfo
  {journal} {Phys. Rev. A}\ }\textbf {\bibinfo {volume} {33}},\ \bibinfo
  {pages} {3742} (\bibinfo {year} {1986})}\BibitemShut {NoStop}%
\bibitem [{\citenamefont {Neese}(2022)}]{Neese2022}%
  \BibitemOpen
  \bibfield  {author} {\bibinfo {author} {\bibfnamefont {F.}~\bibnamefont
  {Neese}},\ }\bibfield  {title} {\bibinfo {title} {{Software update: The ORCA
  program system—Version 5.0}},\ }\href
  {https://doi.org/https://doi.org/10.1002/wcms.1606} {\bibfield  {journal}
  {\bibinfo  {journal} {WIREs Comput. Mol. Sci.}\ ,\ \bibinfo {pages} {e1606}}
  (\bibinfo {year} {2022})}\BibitemShut {NoStop}%
\bibitem [{\citenamefont {Ballhausen}(1962)}]{Ballhausen1962}%
  \BibitemOpen
  \bibfield  {author} {\bibinfo {author} {\bibfnamefont {C.~J.}\ \bibnamefont
  {Ballhausen}},\ }\href@noop {} {\emph {\bibinfo {title} {{Introduction to
  Ligand Field Theory}}}}\ (\bibinfo  {publisher} {McGraw-Hill Book Company
  Inc., US},\ \bibinfo {year} {1962})\BibitemShut {NoStop}%
\bibitem [{\citenamefont {Abragam}\ and\ \citenamefont
  {Bleaney}(2012)}]{AbragamBleaney}%
  \BibitemOpen
  \bibfield  {author} {\bibinfo {author} {\bibfnamefont {A.}~\bibnamefont
  {Abragam}}\ and\ \bibinfo {author} {\bibfnamefont {B.}~\bibnamefont
  {Bleaney}},\ }\href@noop {} {\emph {\bibinfo {title} {Electron paramagnetic
  resonance of transition ions}}}\ (\bibinfo  {publisher} {Oxford University
  Press, Oxford},\ \bibinfo {year} {2012})\BibitemShut {NoStop}%
\bibitem [{\citenamefont {Pilbrow}(1990)}]{Pilbrow1990}%
  \BibitemOpen
  \bibfield  {author} {\bibinfo {author} {\bibfnamefont {J.~R.}\ \bibnamefont
  {Pilbrow}},\ }\href@noop {} {\emph {\bibinfo {title} {{Transition Ion
  Electron Paramagnetic Resonance}}}}\ (\bibinfo  {publisher} {Clarendon Press
  ; Oxford University Press},\ \bibinfo {year} {1990})\BibitemShut {NoStop}%
\bibitem [{\citenamefont {Marts}\ \emph {et~al.}(2017)\citenamefont {Marts},
  \citenamefont {Kaine}, \citenamefont {Baum}, \citenamefont {Clayton},
  \citenamefont {Bennett}, \citenamefont {Cordonnier}, \citenamefont
  {McCarrick}, \citenamefont {Hasheminasab}, \citenamefont {Crandall},
  \citenamefont {Ziegler},\ and\ \citenamefont {Tierney}}]{Marts2017}%
  \BibitemOpen
  \bibfield  {author} {\bibinfo {author} {\bibfnamefont {A.~R.}\ \bibnamefont
  {Marts}}, \bibinfo {author} {\bibfnamefont {J.~C.}\ \bibnamefont {Kaine}},
  \bibinfo {author} {\bibfnamefont {R.~R.}\ \bibnamefont {Baum}}, \bibinfo
  {author} {\bibfnamefont {V.~L.}\ \bibnamefont {Clayton}}, \bibinfo {author}
  {\bibfnamefont {J.~R.}\ \bibnamefont {Bennett}}, \bibinfo {author}
  {\bibfnamefont {L.~J.}\ \bibnamefont {Cordonnier}}, \bibinfo {author}
  {\bibfnamefont {R.}~\bibnamefont {McCarrick}}, \bibinfo {author}
  {\bibfnamefont {A.}~\bibnamefont {Hasheminasab}}, \bibinfo {author}
  {\bibfnamefont {L.~A.}\ \bibnamefont {Crandall}}, \bibinfo {author}
  {\bibfnamefont {C.~J.}\ \bibnamefont {Ziegler}},\ and\ \bibinfo {author}
  {\bibfnamefont {D.~L.}\ \bibnamefont {Tierney}},\ }\bibfield  {title}
  {\bibinfo {title} {{Paramagnetic Resonance of Cobalt(II)
  Trispyrazolylmethanes and Counterion Association}},\ }\href
  {https://doi.org/10.1021/acs.inorgchem.6b02520} {\bibfield  {journal}
  {\bibinfo  {journal} {Inorg. Chem.}\ }\textbf {\bibinfo {volume} {56}},\
  \bibinfo {pages} {618} (\bibinfo {year} {2017})}\BibitemShut {NoStop}%
\bibitem [{\citenamefont {Sugano}\ \emph {et~al.}(1970)\citenamefont {Sugano},
  \citenamefont {Tanabe},\ and\ \citenamefont {Kamimura}}]{Sugano1970}%
  \BibitemOpen
  \bibfield  {author} {\bibinfo {author} {\bibfnamefont {S.}~\bibnamefont
  {Sugano}}, \bibinfo {author} {\bibfnamefont {Y.}~\bibnamefont {Tanabe}},\
  and\ \bibinfo {author} {\bibfnamefont {H.}~\bibnamefont {Kamimura}},\
  }\href@noop {} {\emph {\bibinfo {title} {Multiplets of transition-metal ions
  in crystals}}}\ (\bibinfo  {publisher} {Elsevier, Amsterdam},\ \bibinfo
  {year} {1970})\BibitemShut {NoStop}%
\bibitem [{\citenamefont {Lloret}\ \emph {et~al.}(2008)\citenamefont {Lloret},
  \citenamefont {Julve}, \citenamefont {Cano}, \citenamefont {Ruiz-García},\
  and\ \citenamefont {Pardo}}]{Lloret2008}%
  \BibitemOpen
  \bibfield  {author} {\bibinfo {author} {\bibfnamefont {F.}~\bibnamefont
  {Lloret}}, \bibinfo {author} {\bibfnamefont {M.}~\bibnamefont {Julve}},
  \bibinfo {author} {\bibfnamefont {J.}~\bibnamefont {Cano}}, \bibinfo {author}
  {\bibfnamefont {R.}~\bibnamefont {Ruiz-García}},\ and\ \bibinfo {author}
  {\bibfnamefont {E.}~\bibnamefont {Pardo}},\ }\bibfield  {title} {\bibinfo
  {title} {Magnetic properties of six-coordinated high-spin cobalt({II})
  complexes: Theoretical background and its application},\ }\href
  {https://doi.org/10.1016/j.ica.2008.03.114} {\bibfield  {journal} {\bibinfo
  {journal} {Inorg. Chim. Acta}\ }\textbf {\bibinfo {volume} {361}},\ \bibinfo
  {pages} {3432} (\bibinfo {year} {2008})}\BibitemShut {NoStop}%
\bibitem [{\citenamefont {Murrie}(2010)}]{Murrie2010}%
  \BibitemOpen
  \bibfield  {author} {\bibinfo {author} {\bibfnamefont {M.}~\bibnamefont
  {Murrie}},\ }\bibfield  {title} {\bibinfo {title} {{Cobalt(II)}
  single-molecule magnets},\ }\href {https://doi.org/10.1039/B913279C}
  {\bibfield  {journal} {\bibinfo  {journal} {Chem. Soc. Rev.}\ }\textbf
  {\bibinfo {volume} {39}},\ \bibinfo {pages} {1986} (\bibinfo {year}
  {2010})}\BibitemShut {NoStop}%
\bibitem [{\citenamefont {Atanasov}\ \emph {et~al.}(2015)\citenamefont
  {Atanasov}, \citenamefont {Aravena}, \citenamefont {Suturina}, \citenamefont
  {Bill}, \citenamefont {Maganas},\ and\ \citenamefont {Neese}}]{Atanasov2015}%
  \BibitemOpen
  \bibfield  {author} {\bibinfo {author} {\bibfnamefont {M.}~\bibnamefont
  {Atanasov}}, \bibinfo {author} {\bibfnamefont {D.}~\bibnamefont {Aravena}},
  \bibinfo {author} {\bibfnamefont {E.}~\bibnamefont {Suturina}}, \bibinfo
  {author} {\bibfnamefont {E.}~\bibnamefont {Bill}}, \bibinfo {author}
  {\bibfnamefont {D.}~\bibnamefont {Maganas}},\ and\ \bibinfo {author}
  {\bibfnamefont {F.}~\bibnamefont {Neese}},\ }\bibfield  {title} {\bibinfo
  {title} {First principles approach to the electronic structure, magnetic
  anisotropy and spin relaxation in mononuclear 3d-transition metal single
  molecule magnets},\ }\href {https://doi.org/10.1016/j.ccr.2014.10.015}
  {\bibfield  {journal} {\bibinfo  {journal} {Coord. Chem. Rev.}\ }\textbf
  {\bibinfo {volume} {289-290}},\ \bibinfo {pages} {177} (\bibinfo {year}
  {2015})}\BibitemShut {NoStop}%
\bibitem [{\citenamefont {Kumar}\ \emph {et~al.}(2020)\citenamefont {Kumar},
  \citenamefont {SantaLucia}, \citenamefont {Kaniewska-Laskowska},
  \citenamefont {Lindeman}, \citenamefont {Ozarowski}, \citenamefont
  {Krzystek}, \citenamefont {Ozerov}, \citenamefont {Telser}, \citenamefont
  {Berry},\ and\ \citenamefont {Fiedler}}]{Kumar2020}%
  \BibitemOpen
  \bibfield  {author} {\bibinfo {author} {\bibfnamefont {P.}~\bibnamefont
  {Kumar}}, \bibinfo {author} {\bibfnamefont {D.~J.}\ \bibnamefont
  {SantaLucia}}, \bibinfo {author} {\bibfnamefont {K.}~\bibnamefont
  {Kaniewska-Laskowska}}, \bibinfo {author} {\bibfnamefont {S.~V.}\
  \bibnamefont {Lindeman}}, \bibinfo {author} {\bibfnamefont {A.}~\bibnamefont
  {Ozarowski}}, \bibinfo {author} {\bibfnamefont {J.}~\bibnamefont {Krzystek}},
  \bibinfo {author} {\bibfnamefont {M.}~\bibnamefont {Ozerov}}, \bibinfo
  {author} {\bibfnamefont {J.}~\bibnamefont {Telser}}, \bibinfo {author}
  {\bibfnamefont {J.~F.}\ \bibnamefont {Berry}},\ and\ \bibinfo {author}
  {\bibfnamefont {A.~T.}\ \bibnamefont {Fiedler}},\ }\bibfield  {title}
  {\bibinfo {title} {Probing the magnetic anisotropy of {Co(II)} complexes
  featuring redox-active ligands},\ }\href
  {https://doi.org/10.1021/acs.inorgchem.0c01812} {\bibfield  {journal}
  {\bibinfo  {journal} {Inorg. Chem.}\ }\textbf {\bibinfo {volume} {59}},\
  \bibinfo {pages} {16178} (\bibinfo {year} {2020})}\BibitemShut {NoStop}%
\bibitem [{\citenamefont {Chibotaru}\ and\ \citenamefont
  {Ungur}(2012)}]{Chibotaru2012}%
  \BibitemOpen
  \bibfield  {author} {\bibinfo {author} {\bibfnamefont {L.~F.}\ \bibnamefont
  {Chibotaru}}\ and\ \bibinfo {author} {\bibfnamefont {L.}~\bibnamefont
  {Ungur}},\ }\bibfield  {title} {\bibinfo {title} {Ab initio calculation of
  anisotropic magnetic properties of complexes. {I.} unique definition of
  pseudospin hamiltonians and their derivation},\ }\href
  {https://doi.org/10.1063/1.4739763} {\bibfield  {journal} {\bibinfo
  {journal} {J. Chem. Phys.}\ }\textbf {\bibinfo {volume} {137}},\ \bibinfo
  {pages} {064112} (\bibinfo {year} {2012})}\BibitemShut {NoStop}%
\bibitem [{\citenamefont {Bogdanov}\ \emph {et~al.}(2007)\citenamefont
  {Bogdanov}, \citenamefont {Zhuravlev},\ and\ \citenamefont
  {R\"o\ss{}ler}}]{Bogdanov2007}%
  \BibitemOpen
  \bibfield  {author} {\bibinfo {author} {\bibfnamefont {A.~N.}\ \bibnamefont
  {Bogdanov}}, \bibinfo {author} {\bibfnamefont {A.~V.}\ \bibnamefont
  {Zhuravlev}},\ and\ \bibinfo {author} {\bibfnamefont {U.~K.}\ \bibnamefont
  {R\"o\ss{}ler}},\ }\bibfield  {title} {\bibinfo {title} {{Spin-flop
  transition in uniaxial antiferromagnets: Magnetic phases, reorientation
  effects, and multidomain states}},\ }\href
  {https://doi.org/10.1103/PhysRevB.75.094425} {\bibfield  {journal} {\bibinfo
  {journal} {Phys. Rev. B}\ }\textbf {\bibinfo {volume} {75}},\ \bibinfo
  {pages} {094425} (\bibinfo {year} {2007})}\BibitemShut {NoStop}%
\bibitem [{\citenamefont {Chowki}\ \emph {et~al.}(2016)\citenamefont {Chowki},
  \citenamefont {Kumar}, \citenamefont {Mohapatra},\ and\ \citenamefont
  {Mahajan}}]{Chowki2016}%
  \BibitemOpen
  \bibfield  {author} {\bibinfo {author} {\bibfnamefont {S.}~\bibnamefont
  {Chowki}}, \bibinfo {author} {\bibfnamefont {R.}~\bibnamefont {Kumar}},
  \bibinfo {author} {\bibfnamefont {N.}~\bibnamefont {Mohapatra}},\ and\
  \bibinfo {author} {\bibfnamefont {A.~V.}\ \bibnamefont {Mahajan}},\
  }\bibfield  {title} {\bibinfo {title} {{Long-range antiferromagnetic order
  and possible field induced spin-flop transition in BiMnVO$_5$}},\ }\href
  {https://doi.org/10.1088/0953-8984/28/48/486002} {\bibfield  {journal}
  {\bibinfo  {journal} {J. Phys.: Condens. Matter}\ }\textbf {\bibinfo {volume}
  {28}},\ \bibinfo {pages} {486002} (\bibinfo {year} {2016})}\BibitemShut
  {NoStop}%
\bibitem [{\citenamefont {Liu}\ \emph {et~al.}(2022)\citenamefont {Liu},
  \citenamefont {Ouyang}, \citenamefont {Xiao}, \citenamefont {Cao},
  \citenamefont {Wang}, \citenamefont {Xia}, \citenamefont {He},\ and\
  \citenamefont {Tong}}]{Liu2022}%
  \BibitemOpen
  \bibfield  {author} {\bibinfo {author} {\bibfnamefont {X.~C.}\ \bibnamefont
  {Liu}}, \bibinfo {author} {\bibfnamefont {Z.~W.}\ \bibnamefont {Ouyang}},
  \bibinfo {author} {\bibfnamefont {T.~T.}\ \bibnamefont {Xiao}}, \bibinfo
  {author} {\bibfnamefont {J.~J.}\ \bibnamefont {Cao}}, \bibinfo {author}
  {\bibfnamefont {Z.~X.}\ \bibnamefont {Wang}}, \bibinfo {author}
  {\bibfnamefont {Z.~C.}\ \bibnamefont {Xia}}, \bibinfo {author} {\bibfnamefont
  {Z.~Z.}\ \bibnamefont {He}},\ and\ \bibinfo {author} {\bibfnamefont
  {W.}~\bibnamefont {Tong}},\ }\bibfield  {title} {\bibinfo {title} {{Magnetism
  and ESR of the ${S}_{\text{eff}}=\frac{1}{2}$ antiferromagnet
  ${\mathrm{BaCo}}_{2}{({\mathrm{SeO}}_{3})}_{3}\ifmmode\cdot\else\textperiodcentered\fi{}3{\mathrm{H}}_{2}\mathrm{O}$
  with dimer-chain structure}},\ }\href
  {https://doi.org/10.1103/PhysRevB.105.134417} {\bibfield  {journal} {\bibinfo
   {journal} {Phys. Rev. B}\ }\textbf {\bibinfo {volume} {105}},\ \bibinfo
  {pages} {134417} (\bibinfo {year} {2022})}\BibitemShut {NoStop}%
\bibitem [{\citenamefont {Wang}\ \emph {et~al.}(2017)\citenamefont {Wang},
  \citenamefont {Reschke}, \citenamefont {H\"uvonen}, \citenamefont {Do},
  \citenamefont {Choi}, \citenamefont {Gensch}, \citenamefont {Nagel},
  \citenamefont {R\~o\ om},\ and\ \citenamefont {Loidl}}]{Wang2017}%
  \BibitemOpen
  \bibfield  {author} {\bibinfo {author} {\bibfnamefont {Z.}~\bibnamefont
  {Wang}}, \bibinfo {author} {\bibfnamefont {S.}~\bibnamefont {Reschke}},
  \bibinfo {author} {\bibfnamefont {D.}~\bibnamefont {H\"uvonen}}, \bibinfo
  {author} {\bibfnamefont {S.-H.}\ \bibnamefont {Do}}, \bibinfo {author}
  {\bibfnamefont {K.-Y.}\ \bibnamefont {Choi}}, \bibinfo {author}
  {\bibfnamefont {M.}~\bibnamefont {Gensch}}, \bibinfo {author} {\bibfnamefont
  {U.}~\bibnamefont {Nagel}}, \bibinfo {author} {\bibfnamefont
  {T.}~\bibnamefont {R\~o\ om}},\ and\ \bibinfo {author} {\bibfnamefont
  {A.}~\bibnamefont {Loidl}},\ }\bibfield  {title} {\bibinfo {title} {{Magnetic
  Excitations and Continuum of a Possibly Field-Induced Quantum Spin Liquid in
  $\ensuremath{\alpha}\text{\ensuremath{-}}{\mathrm{RuCl}}_{3}$}},\ }\href
  {https://doi.org/10.1103/PhysRevLett.119.227202} {\bibfield  {journal}
  {\bibinfo  {journal} {Phys. Rev. Lett.}\ }\textbf {\bibinfo {volume} {119}},\
  \bibinfo {pages} {227202} (\bibinfo {year} {2017})}\BibitemShut {NoStop}%
\bibitem [{\citenamefont {Ponomaryov}\ \emph {et~al.}(2017)\citenamefont
  {Ponomaryov}, \citenamefont {Schulze}, \citenamefont {Wosnitza},
  \citenamefont {Lampen-Kelley}, \citenamefont {Banerjee}, \citenamefont {Yan},
  \citenamefont {Bridges}, \citenamefont {Mandrus}, \citenamefont {Nagler},
  \citenamefont {Kolezhuk},\ and\ \citenamefont {Zvyagin}}]{Ponomaryov2017}%
  \BibitemOpen
  \bibfield  {author} {\bibinfo {author} {\bibfnamefont {A.~N.}\ \bibnamefont
  {Ponomaryov}}, \bibinfo {author} {\bibfnamefont {E.}~\bibnamefont {Schulze}},
  \bibinfo {author} {\bibfnamefont {J.}~\bibnamefont {Wosnitza}}, \bibinfo
  {author} {\bibfnamefont {P.}~\bibnamefont {Lampen-Kelley}}, \bibinfo {author}
  {\bibfnamefont {A.}~\bibnamefont {Banerjee}}, \bibinfo {author}
  {\bibfnamefont {J.-Q.}\ \bibnamefont {Yan}}, \bibinfo {author} {\bibfnamefont
  {C.~A.}\ \bibnamefont {Bridges}}, \bibinfo {author} {\bibfnamefont {D.~G.}\
  \bibnamefont {Mandrus}}, \bibinfo {author} {\bibfnamefont {S.~E.}\
  \bibnamefont {Nagler}}, \bibinfo {author} {\bibfnamefont {A.~K.}\
  \bibnamefont {Kolezhuk}},\ and\ \bibinfo {author} {\bibfnamefont {S.~A.}\
  \bibnamefont {Zvyagin}},\ }\bibfield  {title} {\bibinfo {title}
  {{Unconventional spin dynamics in the honeycomb-lattice material
  $\ensuremath{\alpha}\text{\ensuremath{-}}{\mathrm{RuCl}}_{3}$: High-field
  electron spin resonance studies}},\ }\href
  {https://doi.org/10.1103/PhysRevB.96.241107} {\bibfield  {journal} {\bibinfo
  {journal} {Phys. Rev. B}\ }\textbf {\bibinfo {volume} {96}},\ \bibinfo
  {pages} {241107} (\bibinfo {year} {2017})}\BibitemShut {NoStop}%
\bibitem [{\citenamefont {Wellm}\ \emph {et~al.}(2018)\citenamefont {Wellm},
  \citenamefont {Zeisner}, \citenamefont {Alfonsov}, \citenamefont {Wolter},
  \citenamefont {Roslova}, \citenamefont {Isaeva}, \citenamefont {Doert},
  \citenamefont {Vojta}, \citenamefont {B\"uchner},\ and\ \citenamefont
  {Kataev}}]{Wellm2018}%
  \BibitemOpen
  \bibfield  {author} {\bibinfo {author} {\bibfnamefont {C.}~\bibnamefont
  {Wellm}}, \bibinfo {author} {\bibfnamefont {J.}~\bibnamefont {Zeisner}},
  \bibinfo {author} {\bibfnamefont {A.}~\bibnamefont {Alfonsov}}, \bibinfo
  {author} {\bibfnamefont {A.~U.~B.}\ \bibnamefont {Wolter}}, \bibinfo {author}
  {\bibfnamefont {M.}~\bibnamefont {Roslova}}, \bibinfo {author} {\bibfnamefont
  {A.}~\bibnamefont {Isaeva}}, \bibinfo {author} {\bibfnamefont
  {T.}~\bibnamefont {Doert}}, \bibinfo {author} {\bibfnamefont
  {M.}~\bibnamefont {Vojta}}, \bibinfo {author} {\bibfnamefont
  {B.}~\bibnamefont {B\"uchner}},\ and\ \bibinfo {author} {\bibfnamefont
  {V.}~\bibnamefont {Kataev}},\ }\bibfield  {title} {\bibinfo {title}
  {{Signatures of low-energy fractionalized excitations in
  $\ensuremath{\alpha}\text{\ensuremath{-}}{\mathrm{RuCl}}_{3}$ from
  field-dependent microwave absorption}},\ }\href
  {https://doi.org/10.1103/PhysRevB.98.184408} {\bibfield  {journal} {\bibinfo
  {journal} {Phys. Rev. B}\ }\textbf {\bibinfo {volume} {98}},\ \bibinfo
  {pages} {184408} (\bibinfo {year} {2018})}\BibitemShut {NoStop}%
\bibitem [{\citenamefont {Ponomaryov}\ \emph {et~al.}(2020)\citenamefont
  {Ponomaryov}, \citenamefont {Zviagina}, \citenamefont {Wosnitza},
  \citenamefont {Lampen-Kelley}, \citenamefont {Banerjee}, \citenamefont {Yan},
  \citenamefont {Bridges}, \citenamefont {Mandrus}, \citenamefont {Nagler},\
  and\ \citenamefont {Zvyagin}}]{Ponomaryov2020}%
  \BibitemOpen
  \bibfield  {author} {\bibinfo {author} {\bibfnamefont {A.~N.}\ \bibnamefont
  {Ponomaryov}}, \bibinfo {author} {\bibfnamefont {L.}~\bibnamefont
  {Zviagina}}, \bibinfo {author} {\bibfnamefont {J.}~\bibnamefont {Wosnitza}},
  \bibinfo {author} {\bibfnamefont {P.}~\bibnamefont {Lampen-Kelley}}, \bibinfo
  {author} {\bibfnamefont {A.}~\bibnamefont {Banerjee}}, \bibinfo {author}
  {\bibfnamefont {J.-Q.}\ \bibnamefont {Yan}}, \bibinfo {author} {\bibfnamefont
  {C.~A.}\ \bibnamefont {Bridges}}, \bibinfo {author} {\bibfnamefont {D.~G.}\
  \bibnamefont {Mandrus}}, \bibinfo {author} {\bibfnamefont {S.~E.}\
  \bibnamefont {Nagler}},\ and\ \bibinfo {author} {\bibfnamefont {S.~A.}\
  \bibnamefont {Zvyagin}},\ }\bibfield  {title} {\bibinfo {title} {{Nature of
  Magnetic Excitations in the High-Field Phase of
  $\ensuremath{\alpha}\text{\ensuremath{-}}{\mathrm{RuCl}}_{3}$}},\ }\href
  {https://doi.org/10.1103/PhysRevLett.125.037202} {\bibfield  {journal}
  {\bibinfo  {journal} {Phys. Rev. Lett.}\ }\textbf {\bibinfo {volume} {125}},\
  \bibinfo {pages} {037202} (\bibinfo {year} {2020})}\BibitemShut {NoStop}%
\bibitem [{\citenamefont {Sahasrabudhe}\ \emph {et~al.}(2020)\citenamefont
  {Sahasrabudhe}, \citenamefont {Kaib}, \citenamefont {Reschke}, \citenamefont
  {German}, \citenamefont {Koethe}, \citenamefont {Buhot}, \citenamefont
  {Kamenskyi}, \citenamefont {Hickey}, \citenamefont {Becker}, \citenamefont
  {Tsurkan}, \citenamefont {Loidl}, \citenamefont {Do}, \citenamefont {Choi},
  \citenamefont {Gr\"uninger}, \citenamefont {Winter}, \citenamefont {Wang},
  \citenamefont {Valent\'{\i}},\ and\ \citenamefont {van
  Loosdrecht}}]{Sahasrabudhe2020}%
  \BibitemOpen
  \bibfield  {author} {\bibinfo {author} {\bibfnamefont {A.}~\bibnamefont
  {Sahasrabudhe}}, \bibinfo {author} {\bibfnamefont {D.~A.~S.}\ \bibnamefont
  {Kaib}}, \bibinfo {author} {\bibfnamefont {S.}~\bibnamefont {Reschke}},
  \bibinfo {author} {\bibfnamefont {R.}~\bibnamefont {German}}, \bibinfo
  {author} {\bibfnamefont {T.~C.}\ \bibnamefont {Koethe}}, \bibinfo {author}
  {\bibfnamefont {J.}~\bibnamefont {Buhot}}, \bibinfo {author} {\bibfnamefont
  {D.}~\bibnamefont {Kamenskyi}}, \bibinfo {author} {\bibfnamefont
  {C.}~\bibnamefont {Hickey}}, \bibinfo {author} {\bibfnamefont
  {P.}~\bibnamefont {Becker}}, \bibinfo {author} {\bibfnamefont
  {V.}~\bibnamefont {Tsurkan}}, \bibinfo {author} {\bibfnamefont
  {A.}~\bibnamefont {Loidl}}, \bibinfo {author} {\bibfnamefont {S.~H.}\
  \bibnamefont {Do}}, \bibinfo {author} {\bibfnamefont {K.~Y.}\ \bibnamefont
  {Choi}}, \bibinfo {author} {\bibfnamefont {M.}~\bibnamefont {Gr\"uninger}},
  \bibinfo {author} {\bibfnamefont {S.~M.}\ \bibnamefont {Winter}}, \bibinfo
  {author} {\bibfnamefont {Z.}~\bibnamefont {Wang}}, \bibinfo {author}
  {\bibfnamefont {R.}~\bibnamefont {Valent\'{\i}}},\ and\ \bibinfo {author}
  {\bibfnamefont {P.~H.~M.}\ \bibnamefont {van Loosdrecht}},\ }\bibfield
  {title} {\bibinfo {title} {{High-field quantum disordered state in
  $\ensuremath{\alpha}\ensuremath{-}{\mathrm{RuCl}}_{3}$: Spin flips, bound
  states, and multiparticle continuum}},\ }\href
  {https://doi.org/10.1103/PhysRevB.101.140410} {\bibfield  {journal} {\bibinfo
   {journal} {Phys. Rev. B}\ }\textbf {\bibinfo {volume} {101}},\ \bibinfo
  {pages} {140410} (\bibinfo {year} {2020})}\BibitemShut {NoStop}%
\bibitem [{\citenamefont {Winter}\ \emph {et~al.}(2018)\citenamefont {Winter},
  \citenamefont {Riedl}, \citenamefont {Kaib}, \citenamefont {Coldea},\ and\
  \citenamefont {Valent\'{\i}}}]{Winter2018}%
  \BibitemOpen
  \bibfield  {author} {\bibinfo {author} {\bibfnamefont {S.~M.}\ \bibnamefont
  {Winter}}, \bibinfo {author} {\bibfnamefont {K.}~\bibnamefont {Riedl}},
  \bibinfo {author} {\bibfnamefont {D.}~\bibnamefont {Kaib}}, \bibinfo {author}
  {\bibfnamefont {R.}~\bibnamefont {Coldea}},\ and\ \bibinfo {author}
  {\bibfnamefont {R.}~\bibnamefont {Valent\'{\i}}},\ }\bibfield  {title}
  {\bibinfo {title} {{Probing
  $\ensuremath{\alpha}\ensuremath{-}{\mathrm{RuCl}}_{3}$ Beyond Magnetic Order:
  Effects of Temperature and Magnetic Field}},\ }\href
  {https://doi.org/10.1103/PhysRevLett.120.077203} {\bibfield  {journal}
  {\bibinfo  {journal} {Phys. Rev. Lett.}\ }\textbf {\bibinfo {volume} {120}},\
  \bibinfo {pages} {077203} (\bibinfo {year} {2018})}\BibitemShut {NoStop}%
\bibitem [{\citenamefont {Yang}\ \emph {et~al.}(2020)\citenamefont {Yang},
  \citenamefont {Nocera},\ and\ \citenamefont {Affleck}}]{Yang2020}%
  \BibitemOpen
  \bibfield  {author} {\bibinfo {author} {\bibfnamefont {W.}~\bibnamefont
  {Yang}}, \bibinfo {author} {\bibfnamefont {A.}~\bibnamefont {Nocera}},\ and\
  \bibinfo {author} {\bibfnamefont {I.}~\bibnamefont {Affleck}},\ }\bibfield
  {title} {\bibinfo {title} {{Comprehensive study of the phase diagram of the
  spin-$\frac{1}{2}$ Kitaev-Heisenberg-Gamma chain}},\ }\href
  {https://doi.org/10.1103/PhysRevResearch.2.033268} {\bibfield  {journal}
  {\bibinfo  {journal} {Phys. Rev. Research}\ }\textbf {\bibinfo {volume}
  {2}},\ \bibinfo {pages} {033268} (\bibinfo {year} {2020})}\BibitemShut
  {NoStop}%
\bibitem [{\citenamefont {Luo}\ \emph {et~al.}(2021)\citenamefont {Luo},
  \citenamefont {Zhao}, \citenamefont {Wang},\ and\ \citenamefont
  {Kee}}]{Luo2021}%
  \BibitemOpen
  \bibfield  {author} {\bibinfo {author} {\bibfnamefont {Q.}~\bibnamefont
  {Luo}}, \bibinfo {author} {\bibfnamefont {J.}~\bibnamefont {Zhao}}, \bibinfo
  {author} {\bibfnamefont {X.}~\bibnamefont {Wang}},\ and\ \bibinfo {author}
  {\bibfnamefont {H.-Y.}\ \bibnamefont {Kee}},\ }\bibfield  {title} {\bibinfo
  {title} {{Unveiling the phase diagram of a bond-alternating
  spin-$\frac{1}{2}$ $K\text{\ensuremath{-}}\mathrm{\ensuremath{\Gamma}}$
  chain}},\ }\href {https://doi.org/10.1103/PhysRevB.103.144423} {\bibfield
  {journal} {\bibinfo  {journal} {Phys. Rev. B}\ }\textbf {\bibinfo {volume}
  {103}},\ \bibinfo {pages} {144423} (\bibinfo {year} {2021})}\BibitemShut
  {NoStop}%
\bibitem [{\citenamefont {Mac\^edo}\ \emph {et~al.}(2022)\citenamefont
  {Mac\^edo}, \citenamefont {Ramos},\ and\ \citenamefont
  {Pereira}}]{Macedo2022}%
  \BibitemOpen
  \bibfield  {author} {\bibinfo {author} {\bibfnamefont {R.~A.}\ \bibnamefont
  {Mac\^edo}}, \bibinfo {author} {\bibfnamefont {F.~B.}\ \bibnamefont
  {Ramos}},\ and\ \bibinfo {author} {\bibfnamefont {R.~G.}\ \bibnamefont
  {Pereira}},\ }\bibfield  {title} {\bibinfo {title} {{Continuous phase
  transition from a chiral spin state to collinear magnetic order in a zigzag
  chain with Kitaev interactions}},\ }\href
  {https://doi.org/10.1103/PhysRevB.105.205144} {\bibfield  {journal} {\bibinfo
   {journal} {Phys. Rev. B}\ }\textbf {\bibinfo {volume} {105}},\ \bibinfo
  {pages} {205144} (\bibinfo {year} {2022})}\BibitemShut {NoStop}%
\bibitem [{\citenamefont {Motome}\ \emph {et~al.}(2020)\citenamefont {Motome},
  \citenamefont {Sano}, \citenamefont {Jang}, \citenamefont {Sugita},\ and\
  \citenamefont {Kato}}]{Motome2020}%
  \BibitemOpen
  \bibfield  {author} {\bibinfo {author} {\bibfnamefont {Y.}~\bibnamefont
  {Motome}}, \bibinfo {author} {\bibfnamefont {R.}~\bibnamefont {Sano}},
  \bibinfo {author} {\bibfnamefont {S.}~\bibnamefont {Jang}}, \bibinfo {author}
  {\bibfnamefont {Y.}~\bibnamefont {Sugita}},\ and\ \bibinfo {author}
  {\bibfnamefont {Y.}~\bibnamefont {Kato}},\ }\bibfield  {title} {\bibinfo
  {title} {{Materials design of Kitaev spin liquids beyond the
  Jackeli-Khaliullin mechanism}},\ }\href
  {https://doi.org/10.1088/1361-648X/ab8525} {\bibfield  {journal} {\bibinfo
  {journal} {J. Phys.: Condens. Matter}\ }\textbf {\bibinfo {volume} {32}},\
  \bibinfo {pages} {404001} (\bibinfo {year} {2020})}\BibitemShut {NoStop}%
\bibitem [{\citenamefont {Mazin}\ \emph {et~al.}(2012)\citenamefont {Mazin},
  \citenamefont {Jeschke}, \citenamefont {Foyevtsova}, \citenamefont
  {Valent\'{\i}},\ and\ \citenamefont {Khomskii}}]{Mazin2012}%
  \BibitemOpen
  \bibfield  {author} {\bibinfo {author} {\bibfnamefont {I.~I.}\ \bibnamefont
  {Mazin}}, \bibinfo {author} {\bibfnamefont {H.~O.}\ \bibnamefont {Jeschke}},
  \bibinfo {author} {\bibfnamefont {K.}~\bibnamefont {Foyevtsova}}, \bibinfo
  {author} {\bibfnamefont {R.}~\bibnamefont {Valent\'{\i}}},\ and\ \bibinfo
  {author} {\bibfnamefont {D.~I.}\ \bibnamefont {Khomskii}},\ }\bibfield
  {title} {\bibinfo {title} {{${\mathrm{Na}}_{2}{\mathrm{IrO}}_{3}$ as a
  Molecular Orbital Crystal}},\ }\href
  {https://doi.org/10.1103/PhysRevLett.109.197201} {\bibfield  {journal}
  {\bibinfo  {journal} {Phys. Rev. Lett.}\ }\textbf {\bibinfo {volume} {109}},\
  \bibinfo {pages} {197201} (\bibinfo {year} {2012})}\BibitemShut {NoStop}%
\bibitem [{\citenamefont {Komleva}\ \emph {et~al.}(2020)\citenamefont
  {Komleva}, \citenamefont {Khomskii},\ and\ \citenamefont
  {Streltsov}}]{Komleva2020}%
  \BibitemOpen
  \bibfield  {author} {\bibinfo {author} {\bibfnamefont {E.~V.}\ \bibnamefont
  {Komleva}}, \bibinfo {author} {\bibfnamefont {D.~I.}\ \bibnamefont
  {Khomskii}},\ and\ \bibinfo {author} {\bibfnamefont {S.~V.}\ \bibnamefont
  {Streltsov}},\ }\bibfield  {title} {\bibinfo {title} {{Three-site
  transition-metal clusters: Going from localized electrons to molecular
  orbitals}},\ }\href {https://doi.org/10.1103/PhysRevB.102.174448} {\bibfield
  {journal} {\bibinfo  {journal} {Phys. Rev. B}\ }\textbf {\bibinfo {volume}
  {102}},\ \bibinfo {pages} {174448} (\bibinfo {year} {2020})}\BibitemShut
  {NoStop}%
\bibitem [{\citenamefont {Tsirlin}\ and\ \citenamefont
  {Gegenwart}(2022)}]{Tsirlin2022}%
  \BibitemOpen
  \bibfield  {author} {\bibinfo {author} {\bibfnamefont {A.~A.}\ \bibnamefont
  {Tsirlin}}\ and\ \bibinfo {author} {\bibfnamefont {P.}~\bibnamefont
  {Gegenwart}},\ }\bibfield  {title} {\bibinfo {title} {{Kitaev Magnetism
  through the Prism of Lithium Iridate}},\ }\href
  {https://doi.org/10.1002/pssb.202100146} {\bibfield  {journal} {\bibinfo
  {journal} {Phys. Status Solidi B}\ }\textbf {\bibinfo {volume} {259}},\
  \bibinfo {pages} {2100146} (\bibinfo {year} {2022})}\BibitemShut {NoStop}%
\end{thebibliography}
